\documentclass[aps,prl,twocolumn,showpacs,superscriptaddress,longbibliography]{revtex4-2}
\usepackage{amsfonts}
\usepackage{amsmath}
\usepackage{graphicx}
\usepackage{bm}
\usepackage{amssymb}
\usepackage{dcolumn}
\usepackage{color}
\usepackage{soul}
\usepackage{siunitx}
\usepackage{multirow}
\usepackage{booktabs}
\usepackage{diagbox}
\usepackage{verbatim}

\usepackage[colorlinks,
    linkcolor=blue,
    anchorcolor=blue,
    citecolor=blue,
    urlcolor=blue,
]{hyperref}

\begin{document}

    \title{Diameter-Controlled High-Order Vortex States and Magnon Hybridization in VSe$_2$ Nanotubes}

    \author{Jia-Wen Li}
    \email{lijw2025@nanoctr.cn}
    \affiliation{Laboratory of Theoretical and Computational Nanoscience, National Center for Nanoscience and Technology, Chinese Academy of Sciences, Beijing 100190, China}
    \affiliation{Kavli Institute for Theoretical Sciences, University of Chinese Academy of Sciences, Beijing 100049, China}
    
    \author{Xin-Wei Yi}
    \affiliation{Institute of Theoretical Physics, Chinese Academy of Sciences, Beijing 100190, China}
	
	\author{Jin Zhang}
	\affiliation{Laboratory of Theoretical and Computational Nanoscience, National Center for Nanoscience and Technology, Chinese Academy of Sciences, Beijing 100190, China}
	
    \author{Gang Su}
    \email{gsu@ucas.ac.cn}
    \affiliation{Institute of Theoretical Physics, Chinese Academy of Sciences, Beijing 100190, China}
    \affiliation{Kavli Institute for Theoretical Sciences, University of Chinese Academy of Sciences, Beijing 100049, China}
    \affiliation{Physical Science Laboratory, Huairou National Comprehensive Science Center, Beijing 101400, China}
    \affiliation{School of Physical Sciences, University of Chinese Academy of Sciences, Beijing 100049, China}

    \author{Bo Gu}
    \email{gubo@ucas.ac.cn}
    \affiliation{Kavli Institute for Theoretical Sciences, University of Chinese Academy of Sciences, Beijing 100049, China}
    \affiliation{Physical Science Laboratory, Huairou National Comprehensive Science Center, Beijing 101400, China}

    \begin{abstract}
   	Curved magnets offer a rich phase diagram and hold great promise for next-generation spintronic technologies. 
   	This study establishes the paramount significance of high-order vortex states (e.g., $3\varphi$ with winding number $n \ge 2$) in $\text{VSe}_2$ nanotubes, which uniquely enable magnonic functionalities fundamentally inaccessible to conventional magnetic systems. These states arise from diameter-dependent competition between the nearest-neighbor ferromagnetic ($J_1$) and longer-range antiferromagnetic ($J_2/J_3$) couplings, as rigorously validated through density-functional theory calculations and Heisenberg modeling of phase diagrams. 
   	Critically, by the Landau-Lifshitz-Gilbert equation, we find that high-order vortex configurations unlock an intrinsic hybridization mechanism governed by strict orbital angular momentum (OAM) selection rules ($\Delta l = \pm 2(n-1)$) -- a process strictly forbidden in fundamental vortices ($n=1$) -- generating complex high-OAM magnons with measurable topological charge. 
   	This is vividly demonstrated in the $3\varphi$ state, where hybridization between $l = -4, 0,$ and $4$ modes produces eight-petal magnon density patterns. 
   	Such states provide an essential platform-free solution for generating high-OAM magnons, wchich is crucial for spin-wave-based information transport.
    These findings establish a predictive theoretical framework for controlling high-order vortex states in curved magnets and highlight VSe$_2$ nanotubes as a promising platform for exploring complex magnetism and developing future magnonic and spintronic devices.
    \end{abstract}
    \maketitle


	\begin{figure*}[hbpt]
		\centering
		\includegraphics[width=1.75\columnwidth]{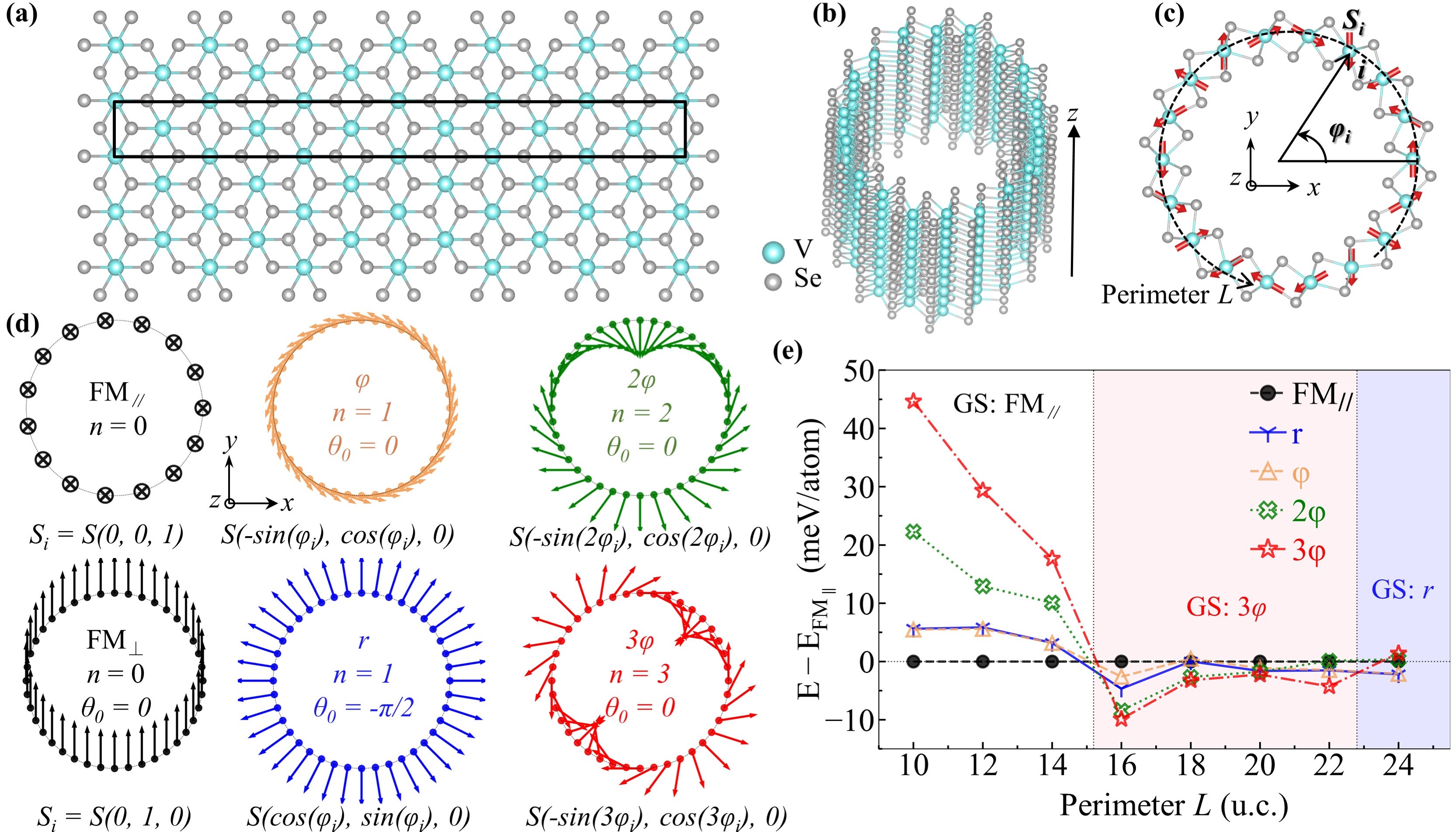}\\
		\caption{
			(a–c) Construction of an armchair VSe$_2$ nanotube from a 1T-VSe$_2$ monolayer (space group $P\bar{3}m1$), showing the rolling direction and resulting cylindrical geometry, where the azimuthal angle of site $i$ is $\varphi_i=2i\pi/L$.
			u.c. denotes unit cell. 
			(d) Magnetic configurations considered in this work: ferromagnetic (FM$_\parallel$, FM$_\perp$), radial ($r$), and $n$-th order vortex states ($n\varphi$) with winding number $n$.  
			(e) Relative energies from noncollinear density functional theory (DFT) for various $L$, showing a diameter-dependent ground-state evolution.			
		}\label{Fig1}
	\end{figure*}

	\textcolor{blue}{{\em Introduction}}---
	Noncollinear spin structures with exotic topological textures are of great interest for both fundamental physics and next-generation spintronic technologies \cite{He2021,Yang2021,Wei2023a,Song2025,Sheka2021}.
	In curved magnets, geometry couples directly to spin configurations, stabilizing noncollinear alignments and complex textures absent in flat systems \cite{Gaididei2014,Sheka2020}.
	This curvature–spin coupling has been explored in various geometries, predicting topological magnetism in Möbius strips \cite{Pylypovskyi2015} and diameter-dependent spin textures in CrI$_3$ nanotubes \cite{Edstroem2022}.
	Experiments further show that geometry can dictate magnetic ground states, such as vortex or onion-like configurations in Ni nanotubes \cite{Rueffer2012}, vortex states in CoFeB and permalloy nanotubes \cite{Wyss2017}, and circumferential magnetization curls in cobalt nanotube arrays \cite{Li2008}.
	This geometric effect can also be harnessed to control magnetic structure \cite{Landeros2007,Zhang2021d}.

	One-dimensional nanotubes are ideal systems for probing such geometric effects.
	Since the discovery of carbon nanotubes \cite{Iijima1991}, their atomically precise structures and diverse chiralities have made them model systems for studying curvature-driven phenomena \cite{Wang2018a,CNTbook}.
	Magnetic nanotubes, realized via magnetic nanoparticle incorporation or direct synthesis of ferromagnetic metals and alloys \cite{Han2009,Korneva2005,Lee2006,Son2005,Guo2021a,Giordano2023,Shpaisman2012}, host novel properties such as nonreciprocal magnon dynamics and controlled domain-wall motion \cite{Yan2012,Yang2018,Koerber2022,Gallardo2022,Otalora2016,SalazarCardona2021}, making them promising spintronic elements \cite{Chumak2015}.

	The advent of two-dimensional (2D) van der Waals (vdW) magnets \cite{Huang2017,Gong2017,Achinuq2021,Lee2021,Zhang2021,Xian2022,Wang2024,Chua2021,Li2022b,Seo2020,Zhang2022,Chen2024,Xiao2022,Bonilla2018,Meng2021}, together with advances in synthesizing transition-metal dichalcogenide (TMD) nanotubes \cite{Aftab2022,Shubina2019,Gao2018,Nakanishi2023,Kamaei2020,Qin2017,Qin2018,Zhang2019d,Kim2022}, offers the prospect of constructing intrinsically magnetic nanotubes from magnetic monolayers.
	Such systems combine curvature tunability with the rich spin physics of vdW magnets.
	A theoretical understanding of intrinsically magnetic nanotubes is urgently needed.
	While some density functional theory (DFT) studies have investigated TMD nanotubes \cite{Evarestov2017,Zhao2015,Boelle2021,Edstroem2022,Bhardwaj2021}, a detailed research on their non-collinear magnetic structure is still needed.
	
	In this Letter, we combine DFT and the Heisenberg model to investigate the magnetic ground states of single-wall VSe$_2$ nanotubes.
	Our calculations reveal a diameter-dependent evolution of the spin texture, including a novel high-order vortex magnetic state 3$\varphi$. 
	According to a Heisenberg model, we find that the vortex magnetic states $n\varphi$ with winding number $n =1,2,3,4$, etc can be developed by competition between the nearest ferromagnetic (FM) and the 2nd- and 3rd-nearest antiferromagnetic (AFM) coupling constants.
	The model reproduces the DFT results across a broad range of tube diameters, confirming the robustness of the mechanism.
	Moreover, through the Landau-Lifshitz-Gilbert equation, we found that high-order vortex states enable the hybridization between magnon modes with different orbital angular momentum (OAM). 
	This hybridization requires non-zero magnetic anisotropy energy (MAE) and is allowed only in high-order vortex states with $n>1$ such as the $3\varphi$ state, enabling controlled generation of high-OAM magnons.
	Our results provide a predictive framework for understanding and engineering high-order vortex states in curved magnets, and identify VSe$_2$ nanotubes as an attractive platform for complex spin textures and advanced magnonic applications.

	\textcolor{blue}{{\em High-order vortex magnetic state 3$\varphi$ in VSe$_2$ nanotubes}}---
	The intrinsic curvature arising from the formation of a magnetic nanotube by rolling up a monolayer (Figs. \ref{Fig1}(a, b)) is the central mechanism expected to drive the system into non-trivial magnetic states.
	To explore the consequences of this curvature, we systematically analyze a set of candidate spin configurations, as depicted in Fig. \ref{Fig1}(d).
	These encompass collinear states-including FM configurations with magnetization parallel (FM$_\parallel$) or perpendicular (FM$_\perp$) to the tube axis ($z$).
	Two collinear AFM configurations were considered (cAFM1, cAFM2), as given in Supplemental Material (SM) \cite{SM}.
	Crucially, due to the cylindrical topology and symmetry as shown in Fig. \ref{Fig1}(c), non-collinear vortex states are allowed, which could be expressed as \cite{Mermin1979,Verba2018,Shibata2006,Edstroem2022,Koerber2022}:
	\begin{align}
		&\vec{S}_i^{(n\varphi)} = S\left(-\sin(2\pi ni/L+\theta_0), \cos(2\pi ni/L+\theta_0), 0\right),
		\label{Eq:S}
	\end{align}
	where $S_i$ denotes the magnetic moment of atom $i$, $L$ is the number of magnetic atoms per cell, and $\theta_0$ is the initial phase. 
	The integer winding number $n$ quantifies how many full rotations the spin completes along the circumference:
	\begin{align}
		n = \frac{1}{2\pi S^2} \int_0^{2\pi} \left( \vec{S} \times \frac{\partial \vec{S}}{\partial \varphi} \right)\cdot\vec{z}~d\varphi,
		\label{Eq:n}
	\end{align}
	where $\vec{S}$ is spin vector confined to the $xy$-plane.
	The parameter $n$ can be an integer (0, $\pm$1, $\pm$2, ...).
	The case $n=0$ corresponds to the collinear FM$_\perp$ and FM$_\parallel$ configurations, while all non-zero cases ($n\neq0$) represent non-collinear states.
	Our investigation focuses on several non-collinear states, including the simple states defined by $n=1$ (the fundamental vortex state $\varphi$ with $\theta_0=0$ and the radial state $r$ with $\theta_0=-\pi/2$), as well as high-order vortex states with $n\geq2$ ($2\varphi$, $3\varphi$, and $4\varphi$) with zero $\theta_0$.

	\begin{figure*}[hbpt]
		\centering
		\includegraphics[width=1.65\columnwidth]{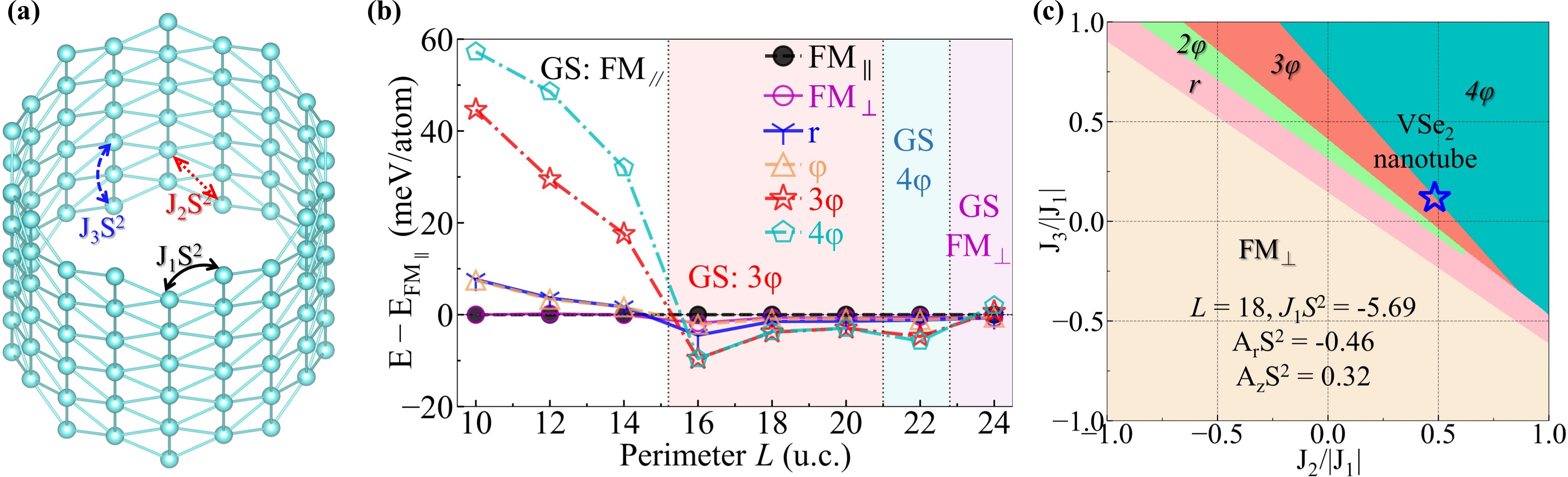}\\
		\caption{
			(a) Schematic of exchange interactions $J_1$, $J_2$, and $J_3$ between V atoms in the cylindrical geometry.  
			(b) Relative energies from the Heisenberg model (Eq. \ref{Eq:E2}) for various perimeter $L$, showing a diameter-dependent ground-state evolution.
			(c) Magnetic phase diagrams for $L=18$, plotted in the $(J_2/|J_1|, J_3/|J_1|)$ parameter space.  
			Regions correspond to stable FM and vortex orders of different winding numbers $n$.  
			The red markers indicate parameters extracted from DFT (Table~\ref{Tab1}), correctly predicting the $3\varphi$ ground state.  
		}\label{Fig2}
	\end{figure*}

	The magnetic nanotubes are constructed from monolayer 1T-VSe$_2$.
	The crystal structure is shown in Fig.~\ref{Fig1}(a) (space group P$\overline{3}$m1, No.~164).
	The calculated in-plane lattice constants are $a_0/\sqrt{3}=b_0=3.5$~\AA, in agreement with experimental value of 3.4~$\pm$~0.1~\AA~\cite{Wong2019}.
	Our calculations show it is a FM metal, consistent with previous studies \cite{Memarzadeh2021,Wong2019,Yu2019,Bonilla2018}. 
	More details see SM \cite{SM}.

	The VSe$_2$ nanotubes are constructed by rolling up the monolayer sheet, as illustrated in Fig. \ref{Fig1}(b).
	We investigate a series of armchair nanotubes with perimeters ($L$), ranging from 10 to 24 repeating number of VSe$_2$ units.
	To verify their ground states, we systematically compared the energies of different magnetic orders (Fig. \ref{Fig1}(d)), computed via DFT.
	The calculations demonstrate a strong dependence of the ground state on perimeter, as summarized in Fig. \ref{Fig1}(e).
	(1) For small perimeters ($L$ = 10, 12, 14), a collinear FM$_\parallel$ configuration is energetically favored.
	(2) For intermediate perimeters ($L$ = 16, 18, 20, 22), the ground state transitions to a novel high-order $3\varphi$ vortex state, as shown in Fig. \ref{Fig1}(c).
	(3) At a bigger perimeter ($L=24$), the radial state ($r$) is the most stable.
	More details see in SM \cite{SM}.

	\begin{table*}[htbp]
		\centering
		\caption{Perimeter, diameter, coupling constants $J_1S^2$, $J_2S^2$ and $J_3S^2$, magnetic anisotropy energy (MAE), Curie temperature ($T\rm_C$), Neel temperature ($T\rm_N$),  the ground state obtained by Eq. \eqref{Eq:H}, and spin-polarized DFT calculation for armchair VSe$_2$ nanotubes with spin-orbital coupling.}
		\label{Tab1}
		\begin{tabular}{c|c|ccc|cc|cc|c}
			\hline\hline
			\makebox[0.10\textwidth][c]{Perimeter}
			&\makebox[0.08\textwidth][c]{Diameter} &\multicolumn{3}{c|}{Coupling constants (meV)}
			&\multicolumn{2}{c|}{MAE (meV)} 
			& \multicolumn{2}{c|}{Ground State}&
			\makebox[0.08\textwidth][c]{
				$T\rm_C$ or $T\rm_N$ }\\
			\cline{3-9}
			(VSe$_2$ units)
			&(nm)
			&\makebox[0.07\textwidth][c]{$J_1S^2$}
			&\makebox[0.066\textwidth][c]{$J_2S^2$}
			&\makebox[0.066\textwidth][c]{$J_3S^2$}
			&\makebox[0.066\textwidth][c]{$A_r S^2$}
			&\makebox[0.066\textwidth][c]{$A_z S^2$}
			&\makebox[0.066\textwidth][c]{Eq. \eqref{Eq:E2}}
			&\makebox[0.066\textwidth][c]{DFT}
			&(K)
			\\\hline
			10 & 1.28 & -16.59& 1.23& -1.86 & 0.14 & 0.07 & FM$_\parallel$ & FM$_\parallel$ &243 ($T\rm_C$)\\
			12 & 1.46 & -16.25& -1.32& 2.11 & 0.24 & -0.06 & FM$_\parallel$ & FM$_\parallel$ &195 ($T\rm_C$)\\
			14 & 1.64 & -14.65& -1.50& 2.71 & 0.01 & -0.04 & FM$_\parallel$ & FM$_\parallel$ &162 ($T\rm_C$)\\
			16 & 1.92 & -5.74& 3.34& 1.47 & -2.02 & -0.54 & 3$\varphi$ & 3$\varphi$ &48 ($T\rm_N$)\\
			18 & 2.12 & -5.69& 2.75& 0.67 & -0.46 & 0.32 & 3$\varphi$ & 3$\varphi$ &43 ($T\rm_N$)\\
			20 & 2.32 & -6.49& 2.97& 0.47 & -0.48 & 0.43 & 3$\varphi$ & 3$\varphi$ &48 ($T\rm_N$)\\
			22 & 2.53 & -8.06& 2.72& 2.11 & -0.07 & 0.40 & 4$\varphi$ & 3$\varphi$ &62 ($T\rm_N$)\\
			24 & 2.73 & -3.80& -0.95& 1.15 & -0.09 & 0.65 & FM$_\perp$ & $r$ &43 ($T\rm_N$)\\
			\hline\hline
		\end{tabular}
	\end{table*}

	\textcolor{blue}{{\em Mechanism of vortex magnetic sates $n\varphi$ with winding number $n \geq 1$ }}---
	To analyze the magnetic properties, we consider a Heisenberg-type Hamiltonian, which can be written as:
	\begin{align}
		H=&\sum\limits_{\left \langle i,j \right \rangle}^{i<j}J_{ij}
		\vec{S_i} \cdot \vec{S_{j}}
		+A_z\sum_{i}S_{i,z}^2
		+A_r\sum_{i}S_{i,r}^2
		+E_0.
		\label{Eq:H}
	\end{align}
	Here, $\vec{S_i}$ and $\vec{S_j}$ are spin operators at magnetic atom sites $i$ and $j$ (Fig. \ref{Fig1}(b)).
	$S_{i,z}$ and $S_{i,r}$ are the projections of $\vec{S}_i$ in $z$ and $r$ directions, respectively.
	$J_{ij}$ are the coupling constants between magnetic atoms, including the first ($J_1$), second ($J_2$) and third nearest ($J_3$) exchange coupling constants, as shown in Fig. \ref{Fig2}(a).
	$A_z$ and $A_r$ are the magnetic anisotropy parameters along the tube axis ($z$) and radial ($r$) directions.
	Energies of different magnetic states derived from this Hamiltonian are:
	\begin{align}
		&E
		=
		\left[1+2\cos\left(2\pi n/L\right)\right]LJ_1S^2
		\nonumber\\&~~
		+\left[2\cos\left(2\pi n/L\right)
		+\cos\left(4\pi n/L\right)\right]LJ_2S^2
		\nonumber\\&~~
		+
		\left[1+2\cos\left(4\pi n/L\right)\right]LJ_3S^2
		+E_{\rm MAE}+E_0,
		\nonumber\\&
		E_{\rm MAE}=\begin{cases}A_zS^2L~~({\rm FM}_\parallel)\\A_rS^2L~~(r)\\0~~(\varphi)\\0.5A_rS^2L~~({\rm FM}_\perp,~2\varphi,~3\varphi,...)\\\end{cases}
		\label{Eq:E2}
	\end{align}
	The energy of a given state is thus determined by the interplay between the structural parameter ($L$) and the magnetic parameters ($J_1$, $J_2$, $J_3$, $A_z$ and $A_r$).
	In addition, it is the MAE parameter that makes the energies of states with the same $n$ unequal, such as $r$ and $\varphi$ states.
	
	The predicted ground states by this Heisenberg model agree well with our DFT results in Fig. \ref{Fig1}(e).
	As shown in Fig. \ref{Fig2}(b), the relative energies of different magnetic states, obtained by substituting values of magnetic parameters from Table \ref{Tab1} into Eq. \eqref{Eq:E2}, exhibit a clear diameter-dependent evolution of the ground state.
	Consistent with the DFT results, the model shows a ground-state transition from the collinear FM$_{\parallel}$ state at small perimeters ($L = 10, 12, 14$) to the non-collinear $3\varphi$ state at intermediate perimeters ($L = 16, 18, 20$).
	At a larger perimeter of $L=22$, the model predicts a $4\varphi$ ground state, followed by a FM$_{\perp}$ state at $L=24$.
	The energy differences between states with the same winding number, such as $E{\rm_{FM_\perp}}-E{\rm_{FM_\parallel}}$ and $E{_\varphi}-E{_r}$, are small as they are governed by the small MAE terms.
	While our Heisenberg model truncated at $J_3$ shows good agreement with DFT for $L=10\sim20$, deviations for larger diameters ($L$ = 22, 24) arise because the energies of competing states become nearly degenerate at this diameter range.
	This proximity in energy makes the ground state sensitive to neglected long-range exchange interactions, leading to the minor discrepancies between the model and DFT calculations.
	Nevertheless, the model successfully captures the essential physics of the high-order vortex formation.

	To visualize how the interplay of exchange parameters determines the ground state, we constructed magnetic phase diagrams in the parameter space of $J_2/|J_1|$ versus $J_3/|J_1|$.
	The phase diagram of VSe$_2$ nanotube with $L=18$ considering magnetic parameters in Table \ref{Tab1} is shown in Fig.~\ref{Fig2}(c).
	The phase diagram is partitioned into distinct regions for the ferromagnetic (FM$_\perp$) state and various non-collinear vortex states.
	The FM state is favored when all couplings are ferromagnetic (the 3rd quadrant) or when the competition is weak with small $J_2$ and $J_3$ relative to $J_1$.
	As shown in Table \ref{Tab1}, while the nearest-neighbor coupling $J_1$ remains ferromagnetic, the longer-range couplings $J_2$
	and $J_3$ vary in sign with the perimeter. 
	This variation introduces a crucial competition that generates sufficient magnetic frustration to stabilize non-collinear vortex states. 
	Consequently, the phase space is partitioned into regions of different vortex orders.
	In addition, it is the MAE that breaks the energy degeneracy between states with the same $n$.
	FM$_\perp$ is more stable than FM$_\parallel$ with $L=18$ (Fig.~\ref{Fig2}), because of the negative value of magnetic anisotropy parameter ($A_r-A_z$).	
	Considering the parameters in Table \ref{Tab1}, our model correctly predicts the $3\varphi$ ground state for $L=18$ nanotube, in agreement with our DFT results.
	This agreement holds well across all studied perimeters, strongly validates our analytical model.

	To further probe the robustness of the competition mechanism, we constructed phase diagrams for more parameters, as detailed in SM~\cite{SM}.
	We found that an AFM $J_1$ results in a more intricate phase diagram, yet the fundamental partitioning of the parameter space by competing interactions remains.
	In addition, our analysis shows that the MAE acts as a small perturbation that does not significantly alter the overall phase diagram structure.
	For non-collinear states with $L=16\sim24$, the negative values of magnetic anisotropy parameter ($A_r-A_z$) confine spins to the $x-y$ plane (Table \ref{Tab1}).
	This positive value energetically penalizes any out-of-plane spin component, ensuring a planar magnetic structure.
	In addition, we confirm no spiral modulation along the $z$ axis, as the magnetic moments in adjacent unit cells are in phase \cite{SM}.

	Our model successfully predicts high-order vortex magnetic states, which are forbidden under the framework of continuum limit in magnetism \cite{aharoni2000introduction}.
	The continuum model is valid when the angles between neighboring spins are small ($n\varphi_0 \ll 1$, or $n/L\ll1$), corresponding to large diameter $L$ and small $n$, leading to an unphysical energy trend that contradicts our DFT results.
	Detailed analysis is supplied in the SM \cite{SM}.

	\textcolor{blue}{{\em Magnons in vortex magnetic states $\varphi$ and $3\varphi$ }}---
	Magnons can carry non-zero orbital angular momentum (OAM) \cite{Jia2019}.
	This OAM is a measurable quantity protected against damping, making it a promising candidate for information transport.
	Magnons with non-zero OAM can be excited through various mechanisms, including the Aharanov-Casher effect \cite{Jia2019}, complex external fields \cite{Jiang2020,Chen2020c}, magnonic spiral phase plates \cite{Jia2019a}, skyrmion-textured domain walls \cite{Lee2022}, and spin-to-orbital angular momentum conversion \cite{Li2022c}.
	The detection of magnon OAM is achievable via spin pumping currents and the inverse spin Hall effect \cite{Chen2020c,Jia2021}.
	However, the generation of magnon modes with large OAM values remains a significant challenge.

	To analyze the magnon (spin-wave) excitations under the $n\varphi$ vortex ground state, we express the magnetization perturbation $\vec{m}_i^{(n)}(t)$ at each lattice site $i$ and time $t$ as a linear combination of two local orthogonal basis vectors: $\vec{m}_i^{(n)}(t)=a_i^{(n)}(t)\vec{p}_i^{(n)}+b_i^{(n)}(t)\vec{z}$, where the in-plane basis vector is defined as $\vec{p}_i^{(n)} = (\cos(n\varphi_i), \sin(n\varphi_i), 0)$.	
	A Fourier transform maps these real-space amplitudes into their reciprocal-space counterparts, $a_{l,k_z}^{(n)}$ and $b_{l,k_z}^{(n)}$, through $a_{l,k_z}^{(n)} = \sum_{i} a_{i}^{(n)}(t) e^{-i(l\varphi_i + k_z z_i - \omega_{l,k_z}^{(n)} t)}/\sqrt{N}$.
	Here, $l$ is the OAM quantum number arising from the periodic boundary condition around the circumference, and it quantifies the magnon's OAM along the tube axis as $L_z = \hbar l$ \cite{Jia2019,Wang2022a}. 
	$\omega_{l,k_z}^{(n)}$ is the frequency.
	$k_z$ is the wave vector along the tube axis ($\vec{z}$).
	For brevity, $t$ and $k_z$ are henceforth suppressed.
	Due to the periodic boundary condition, we set $l\in[-L/2+1, L/2]$.	
	The complete set of modes for the system is described by a $2L$-dimensional state vector, $\Psi^{(n)} = (\dots,a_l^{(n)}, b_l^{(n)},\dots)^T$.	
	Through the Landau-Lifshitz-Gilbert equation \cite{Gilbert2004}, the magnon dynamics are governed by the following equation of motion \cite{SM}:
	\begin{align}\label{eq6}
		i\omega^{(n)}_l
		\begin{pmatrix}
			1&-\alpha \\
			\alpha&1
		\end{pmatrix}
		\begin{pmatrix}
			a_l \\
			b_l
		\end{pmatrix}
		= \hat{\mathbf{A}}_{l,l}^{(n)} 
		\begin{pmatrix} a_l \\ b_l \end{pmatrix} + \hat{\mathbf{A}}_{l,l\pm\Delta l}^{(n)} \begin{pmatrix} a_{l\pm\Delta l} \\ b_{l\pm\Delta l} \end{pmatrix},
	\end{align}
	where $\alpha$ is the phenomenological Gilbert damping parameter and $\omega^{(n)}_l$ are the eigenfrequencies.	
	$\hat{\mathbf{A}}_{l,l}^{(n)}$ represent intra-mode interactions, while $\hat{\mathbf{A}}_{l, l \pm \Delta l}$ describe the inter-mode coupling:
	\begin{align}\label{eq7}
		&\hat{\mathbf{A}}_{l,l}^{(n)} =
		-\frac{\gamma}{\hbar}
		\begin{pmatrix}
			0 & \mathcal{J}_l - \mathcal{J}_n + (2A_z -A_r)S^2 \\
			\mathcal{J}_n - \frac{\mathcal{J}_{l-n} + \mathcal{J}_{l+n}}{2} & 0
		\end{pmatrix},
		\nonumber\\&
		\hat{\mathbf{A}}_{l, l \pm \Delta l}^{(n)} =
		-\frac{\gamma}{\hbar}A_rS^2
		\begin{pmatrix}
			0 & \frac{1}{2} \\
			-1 & 0
		\end{pmatrix}, ~~
		\Delta l=2(n-1),
	\end{align}	
	where $\gamma$ is the gyromagnetic ratio and $\hbar$ is the planck constant.
	Here, $\mathcal{J}_{l}=\sum_{j}J_{ij}S^2e^{-il(\varphi_{i}-\varphi_{j})}$ is the Fourier transform of the exchange interaction constants considering $J_1$, $J_2$ and $J_3$.
	Detailed produces of Eq. \eqref{eq6} and Eq. \eqref{eq7} are given in SM \cite{SM}.

	\begin{figure}[hbpt]
		\centering
		\includegraphics[width=0.99\columnwidth]{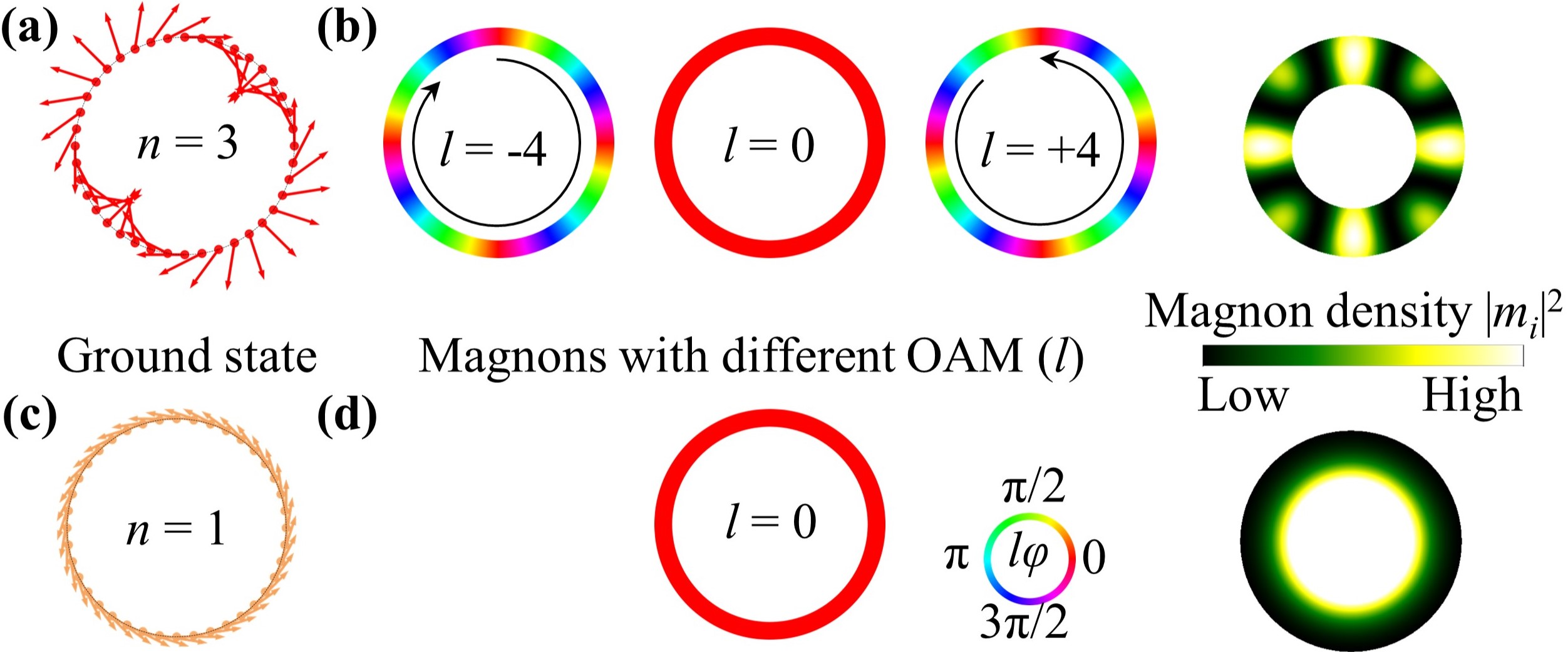}\\
		\caption{
			Orbital angular momentum (OAM) hybridization of magnon modes enabled by high-order vortex states.
			(a, b) In the $n=3$ vortex state 3$\varphi$, magnon modes hybridize according to the selection rule $\Delta l = \pm 2(n-1)=\pm4$, coupling the $l=0$ mode with $l=\pm4$ modes. 
			This creates a hybridized state with a characteristic eight-petal density pattern.
			(c, d) For the $n=1$ state, $\Delta l = 0$ and no hybridization occurs, leaving
			the $l=0$ mode a simple ring.
			The color wheel indicates the magnon phase factor $e^{il\varphi}$.
		}\label{Fig3}
	\end{figure}
		
	The existence of inter-mode interaction requires radial MAE and the ground state as high-order vortex ---$A_r(n-1)\neq0$--- leading to two distinct physical scenarios.	
	For conventional ground states, such as the axial ferromagnetic (FM$_\parallel$) state, or for fundamental ($n=1$) vortex states ($\varphi$ and $r$), the condition is not met, and magnon modes with different $l$ remain decoupled.
	Conversely, in high-order vortex states ($n > 1$) with finite radial anisotropy, the non-zero $\hat{\mathbf{A}}_{l,l\pm\Delta l}^{(n)}$ term induces the formation of hybridized magnon states.
	These eigenstates are linear superpositions of multiple pristine $l$ modes, governed by the specific OAM selection rule $\Delta l = \pm 2(n-1)$, as schematically illustrated for $n=3$ in Fig.~\ref{Fig3}(a).
	The hybridization of $l=0$ with $l=\pm4$ modes, for instance, produces a local magnon density with a flower-like symmetry pattern featuring eight petals \cite{Jia2019}.
	The local magnon density distribution $|m_{i}^{(n)}|^2$, which represents the amplitude of the spin-wave excitation at site $i$, is defined as the squared magnitude of the magnetization perturbation vector as $|\vec{m}_{i}^{(n)}|^2 = (a_{i}^{(n)})^2 + (b_{i}^{(n)})^2$ \cite{Jia2019}.
	In contrast, no such hybridization occurs in the fundamental $\varphi$ ($n=1$) state, as shown in Fig.~\ref{Fig3}(b).
	The resulting spatial patterns can be probed by various experimental techniques, including time-resolved Faraday and Kerr effects, Brillouin light scattering spectroscopy, etc. \cite{Chumak2015}
	Therefore, magnetic nanotubes in a high-order vortex state can serve as an effective intrinsic generators of magnons with large OAM for advanced applications \cite{Chen2020c,Jiang2020,Jia2021}.
	Further details on the procedure are provided in the SM \cite{SM}.

	Furthermore, our results demonstrate that mode hybridization in high-order vortex states provides a new mechanism for opening a magnonic band gap, a key element for spintronic devices \cite{Chumak2015}.
	This mechanism requires a substantial MAE.
	Although the intrinsic MAE of the material is too weak to open a gap, our calculations show that a 30-fold enhancement of the MAE creates a prominent band gap.
	This suggests that magnetic nanotubes engineered to have strong MAE could naturally host such a magnonic gap.
	Detailed calculations are presented in SM \cite{SM}.

    \textcolor{blue}{{\em Conclusion}}---
	We systematically investigated the magnetic ground states of VSe$_2$ nanotubes using DFT and predicted diameter-dependent high-order vortex states.
	By formulating a Heisenberg model including the first ($J_1$), second ($J_2$) and third ($J_3$) nearest magnetic coupling constants, we revealed that these vortex orders originate from the competition between the FM $J_1$ and AFM $J_2$ and $J_3$. 
	The resulting magnetic phase diagrams are in good agreement with our DFT calculations, confirming the robustness of the proposed mechanism.
	Moreover, we demonstrated that high-order vortex states with $n>1$ such as $3\varphi$, in the presence of magnetic anisotropy, enable hybridization of magnon modes with different OAMs, providing an intrinsic route to generate high-OAM magnons.
	These findings establish a predictive theoretical framework for understanding and controlling high-order vortex states in curved magnets, and highlight VSe$_2$ nanotubes as a promising platform for exploring complex spin textures and developing next-generation magnonic and spintronic devices.

    \textcolor{blue}{{\em Acknowledgements}}---
    This work is supported by National Key R\&D Program of China (Grant No. 2022YFA1405100), Chinese Academy of Sciences Project for Young Scientists in Basic Research (Grant No. YSBR-030), and Basic Research Program of the Chinese Academy of Sciences Based on Major Scientific Infrastructures (Grant No. JZHKYPT-
    2021-08). GS was supported in part by the Innovation Program for Quantum Science and Technology under Grant No. 2024ZD0300500, NSFC Nos. 12534009 and 12447101, and the Strategic Priority Research Program of Chinese Academy of Sciences (Grant No. XDB1270000).

    \bibliography{ref1}

\begin{thebibliography}{76}%
\makeatletter
\providecommand \@ifxundefined [1]{%
 \@ifx{#1\undefined}
}%
\providecommand \@ifnum [1]{%
 \ifnum #1\expandafter \@firstoftwo
 \else \expandafter \@secondoftwo
 \fi
}%
\providecommand \@ifx [1]{%
 \ifx #1\expandafter \@firstoftwo
 \else \expandafter \@secondoftwo
 \fi
}%
\providecommand \natexlab [1]{#1}%
\providecommand \enquote  [1]{``#1''}%
\providecommand \bibnamefont  [1]{#1}%
\providecommand \bibfnamefont [1]{#1}%
\providecommand \citenamefont [1]{#1}%
\providecommand \href@noop [0]{\@secondoftwo}%
\providecommand \href [0]{\begingroup \@sanitize@url \@href}%
\providecommand \@href[1]{\@@startlink{#1}\@@href}%
\providecommand \@@href[1]{\endgroup#1\@@endlink}%
\providecommand \@sanitize@url [0]{\catcode `\\12\catcode `\$12\catcode
  `\&12\catcode `\#12\catcode `\^12\catcode `\_12\catcode `\%12\relax}%
\providecommand \@@startlink[1]{}%
\providecommand \@@endlink[0]{}%
\providecommand \url  [0]{\begingroup\@sanitize@url \@url }%
\providecommand \@url [1]{\endgroup\@href {#1}{\urlprefix }}%
\providecommand \urlprefix  [0]{URL }%
\providecommand \Eprint [0]{\href }%
\providecommand \doibase [0]{https://doi.org/}%
\providecommand \selectlanguage [0]{\@gobble}%
\providecommand \bibinfo  [0]{\@secondoftwo}%
\providecommand \bibfield  [0]{\@secondoftwo}%
\providecommand \translation [1]{[#1]}%
\providecommand \BibitemOpen [0]{}%
\providecommand \bibitemStop [0]{}%
\providecommand \bibitemNoStop [0]{.\EOS\space}%
\providecommand \EOS [0]{\spacefactor3000\relax}%
\providecommand \BibitemShut  [1]{\csname bibitem#1\endcsname}%
\let\auto@bib@innerbib\@empty
\bibitem [{\citenamefont {He}\ \emph {et~al.}(2021)\citenamefont {He},
  \citenamefont {Hughes}, \citenamefont {Armitage}, \citenamefont {Tokura},\
  and\ \citenamefont {Wang}}]{He2021}%
  \BibitemOpen
  \bibfield  {author} {\bibinfo {author} {\bibfnamefont {Q.~L.}\ \bibnamefont
  {He}}, \bibinfo {author} {\bibfnamefont {T.~L.}\ \bibnamefont {Hughes}},
  \bibinfo {author} {\bibfnamefont {N.~P.}\ \bibnamefont {Armitage}}, \bibinfo
  {author} {\bibfnamefont {Y.}~\bibnamefont {Tokura}},\ and\ \bibinfo {author}
  {\bibfnamefont {K.~L.}\ \bibnamefont {Wang}},\ }\bibfield  {title} {\bibinfo
  {title} {Topological spintronics and magnetoelectronics},\ }\href
  {https://doi.org/10.1038/s41563-021-01138-5} {\bibfield  {journal} {\bibinfo
  {journal} {Nat. Mater.}\ }\textbf {\bibinfo {volume} {21}},\ \bibinfo {pages}
  {15} (\bibinfo {year} {2021})}\BibitemShut {NoStop}%
\bibitem [{\citenamefont {Yang}\ \emph {et~al.}(2021)\citenamefont {Yang},
  \citenamefont {Naaman}, \citenamefont {Paltiel},\ and\ \citenamefont
  {Parkin}}]{Yang2021}%
  \BibitemOpen
  \bibfield  {author} {\bibinfo {author} {\bibfnamefont {S.-H.}\ \bibnamefont
  {Yang}}, \bibinfo {author} {\bibfnamefont {R.}~\bibnamefont {Naaman}},
  \bibinfo {author} {\bibfnamefont {Y.}~\bibnamefont {Paltiel}},\ and\ \bibinfo
  {author} {\bibfnamefont {S.~S.~P.}\ \bibnamefont {Parkin}},\ }\bibfield
  {title} {\bibinfo {title} {Chiral spintronics},\ }\href
  {https://doi.org/10.1038/s42254-021-00302-9} {\bibfield  {journal} {\bibinfo
  {journal} {Nat. Rev. Phys.}\ }\textbf {\bibinfo {volume} {3}},\ \bibinfo
  {pages} {328} (\bibinfo {year} {2021})}\BibitemShut {NoStop}%
\bibitem [{\citenamefont {Wei}\ \emph {et~al.}(2023)\citenamefont {Wei},
  \citenamefont {Ding}, \citenamefont {Zhang}, \citenamefont {Li},
  \citenamefont {Zeng},\ and\ \citenamefont {Fu}}]{Wei2023a}%
  \BibitemOpen
  \bibfield  {author} {\bibinfo {author} {\bibfnamefont {N.}~\bibnamefont
  {Wei}}, \bibinfo {author} {\bibfnamefont {Y.}~\bibnamefont {Ding}}, \bibinfo
  {author} {\bibfnamefont {J.}~\bibnamefont {Zhang}}, \bibinfo {author}
  {\bibfnamefont {L.}~\bibnamefont {Li}}, \bibinfo {author} {\bibfnamefont
  {M.}~\bibnamefont {Zeng}},\ and\ \bibinfo {author} {\bibfnamefont
  {L.}~\bibnamefont {Fu}},\ }\bibfield  {title} {\bibinfo {title} {Curvature
  geometry in 2\uppercase{D} materials},\ }\href
  {https://doi.org/10.1093/nsr/nwad145} {\bibfield  {journal} {\bibinfo
  {journal} {Natl. Sci. Rev.}\ }\textbf {\bibinfo {volume} {10}},\ \bibinfo
  {pages} {nwad145} (\bibinfo {year} {2023})}\BibitemShut {NoStop}%
\bibitem [{\citenamefont {Song}\ \emph {et~al.}(2025)\citenamefont {Song},
  \citenamefont {Stavrić}, \citenamefont {Barone}, \citenamefont {Droghetti},
  \citenamefont {Antonenko}, \citenamefont {Venderbos}, \citenamefont
  {Occhialini}, \citenamefont {Ilyas}, \citenamefont {Ergeçen}, \citenamefont
  {Gedik}, \citenamefont {Cheong}, \citenamefont {Fernandes}, \citenamefont
  {Picozzi},\ and\ \citenamefont {Comin}}]{Song2025}%
  \BibitemOpen
  \bibfield  {author} {\bibinfo {author} {\bibfnamefont {Q.}~\bibnamefont
  {Song}}, \bibinfo {author} {\bibfnamefont {S.}~\bibnamefont {Stavrić}},
  \bibinfo {author} {\bibfnamefont {P.}~\bibnamefont {Barone}}, \bibinfo
  {author} {\bibfnamefont {A.}~\bibnamefont {Droghetti}}, \bibinfo {author}
  {\bibfnamefont {D.~S.}\ \bibnamefont {Antonenko}}, \bibinfo {author}
  {\bibfnamefont {J.~W.~F.}\ \bibnamefont {Venderbos}}, \bibinfo {author}
  {\bibfnamefont {C.~A.}\ \bibnamefont {Occhialini}}, \bibinfo {author}
  {\bibfnamefont {B.}~\bibnamefont {Ilyas}}, \bibinfo {author} {\bibfnamefont
  {E.}~\bibnamefont {Ergeçen}}, \bibinfo {author} {\bibfnamefont
  {N.}~\bibnamefont {Gedik}}, \bibinfo {author} {\bibfnamefont {S.-W.}\
  \bibnamefont {Cheong}}, \bibinfo {author} {\bibfnamefont {R.~M.}\
  \bibnamefont {Fernandes}}, \bibinfo {author} {\bibfnamefont {S.}~\bibnamefont
  {Picozzi}},\ and\ \bibinfo {author} {\bibfnamefont {R.}~\bibnamefont
  {Comin}},\ }\bibfield  {title} {\bibinfo {title} {Electrical switching of a
  $p$-wave magnet},\ }\href {https://doi.org/10.1038/s41586-025-09034-7}
  {\bibfield  {journal} {\bibinfo  {journal} {Nature}\ }\textbf {\bibinfo
  {volume} {642}},\ \bibinfo {pages} {64} (\bibinfo {year} {2025})}\BibitemShut
  {NoStop}%
\bibitem [{\citenamefont {Sheka}(2021)}]{Sheka2021}%
  \BibitemOpen
  \bibfield  {author} {\bibinfo {author} {\bibfnamefont {D.~D.}\ \bibnamefont
  {Sheka}},\ }\bibfield  {title} {\bibinfo {title} {A perspective on
  curvilinear magnetism},\ }\href {https://doi.org/10.1063/5.0048891}
  {\bibfield  {journal} {\bibinfo  {journal} {Appl. Phys. Lett.}\ }\textbf
  {\bibinfo {volume} {118}},\ \bibinfo {pages} {230502} (\bibinfo {year}
  {2021})}\BibitemShut {NoStop}%
\bibitem [{\citenamefont {Gaididei}\ \emph {et~al.}(2014)\citenamefont
  {Gaididei}, \citenamefont {Kravchuk},\ and\ \citenamefont
  {Sheka}}]{Gaididei2014}%
  \BibitemOpen
  \bibfield  {author} {\bibinfo {author} {\bibfnamefont {Y.}~\bibnamefont
  {Gaididei}}, \bibinfo {author} {\bibfnamefont {V.~P.}\ \bibnamefont
  {Kravchuk}},\ and\ \bibinfo {author} {\bibfnamefont {D.~D.}\ \bibnamefont
  {Sheka}},\ }\bibfield  {title} {\bibinfo {title} {Curvature effects in thin
  magnetic shells},\ }\href {https://doi.org/10.1103/physrevlett.112.257203}
  {\bibfield  {journal} {\bibinfo  {journal} {Phys. Rev. Lett.}\ }\textbf
  {\bibinfo {volume} {112}},\ \bibinfo {pages} {257203} (\bibinfo {year}
  {2014})}\BibitemShut {NoStop}%
\bibitem [{\citenamefont {Sheka}\ \emph {et~al.}(2020)\citenamefont {Sheka},
  \citenamefont {Pylypovskyi}, \citenamefont {Landeros}, \citenamefont
  {Gaididei}, \citenamefont {Kákay},\ and\ \citenamefont
  {Makarov}}]{Sheka2020}%
  \BibitemOpen
  \bibfield  {author} {\bibinfo {author} {\bibfnamefont {D.~D.}\ \bibnamefont
  {Sheka}}, \bibinfo {author} {\bibfnamefont {O.~V.}\ \bibnamefont
  {Pylypovskyi}}, \bibinfo {author} {\bibfnamefont {P.}~\bibnamefont
  {Landeros}}, \bibinfo {author} {\bibfnamefont {Y.}~\bibnamefont {Gaididei}},
  \bibinfo {author} {\bibfnamefont {A.}~\bibnamefont {Kákay}},\ and\ \bibinfo
  {author} {\bibfnamefont {D.}~\bibnamefont {Makarov}},\ }\bibfield  {title}
  {\bibinfo {title} {Nonlocal chiral symmetry breaking in curvilinear magnetic
  shells},\ }\href {https://doi.org/10.1038/s42005-020-0387-2} {\bibfield
  {journal} {\bibinfo  {journal} {Commun. Phys.}\ }\textbf {\bibinfo {volume}
  {3}},\ \bibinfo {pages} {128} (\bibinfo {year} {2020})}\BibitemShut {NoStop}%
\bibitem [{\citenamefont {Pylypovskyi}\ \emph {et~al.}(2015)\citenamefont
  {Pylypovskyi}, \citenamefont {Kravchuk}, \citenamefont {Sheka}, \citenamefont
  {Makarov}, \citenamefont {Schmidt},\ and\ \citenamefont
  {Gaididei}}]{Pylypovskyi2015}%
  \BibitemOpen
  \bibfield  {author} {\bibinfo {author} {\bibfnamefont {O.~V.}\ \bibnamefont
  {Pylypovskyi}}, \bibinfo {author} {\bibfnamefont {V.~P.}\ \bibnamefont
  {Kravchuk}}, \bibinfo {author} {\bibfnamefont {D.~D.}\ \bibnamefont {Sheka}},
  \bibinfo {author} {\bibfnamefont {D.}~\bibnamefont {Makarov}}, \bibinfo
  {author} {\bibfnamefont {O.~G.}\ \bibnamefont {Schmidt}},\ and\ \bibinfo
  {author} {\bibfnamefont {Y.}~\bibnamefont {Gaididei}},\ }\bibfield  {title}
  {\bibinfo {title} {Coupling of chiralities in spin and physical spaces: The
  \uppercase{M}öbius ring as a case study},\ }\href
  {https://doi.org/10.1103/physrevlett.114.197204} {\bibfield  {journal}
  {\bibinfo  {journal} {Phys. Rev. Lett.}\ }\textbf {\bibinfo {volume} {114}},\
  \bibinfo {pages} {197204} (\bibinfo {year} {2015})}\BibitemShut {NoStop}%
\bibitem [{\citenamefont {Edström}\ \emph {et~al.}(2022)\citenamefont
  {Edström}, \citenamefont {Amoroso}, \citenamefont {Picozzi}, \citenamefont
  {Barone},\ and\ \citenamefont {Stengel}}]{Edstroem2022}%
  \BibitemOpen
  \bibfield  {author} {\bibinfo {author} {\bibfnamefont {A.}~\bibnamefont
  {Edström}}, \bibinfo {author} {\bibfnamefont {D.}~\bibnamefont {Amoroso}},
  \bibinfo {author} {\bibfnamefont {S.}~\bibnamefont {Picozzi}}, \bibinfo
  {author} {\bibfnamefont {P.}~\bibnamefont {Barone}},\ and\ \bibinfo {author}
  {\bibfnamefont {M.}~\bibnamefont {Stengel}},\ }\bibfield  {title} {\bibinfo
  {title} {Curved magnetism in \uppercase{C}r\uppercase{I}$_3$},\ }\href
  {https://doi.org/10.1103/physrevlett.128.177202} {\bibfield  {journal}
  {\bibinfo  {journal} {Phys. Rev. Lett.}\ }\textbf {\bibinfo {volume} {128}},\
  \bibinfo {pages} {177202} (\bibinfo {year} {2022})}\BibitemShut {NoStop}%
\bibitem [{\citenamefont {Rüffer}\ \emph {et~al.}(2012)\citenamefont
  {Rüffer}, \citenamefont {Huber}, \citenamefont {Berberich}, \citenamefont
  {Albert}, \citenamefont {Russo-Averchi}, \citenamefont {Heiss}, \citenamefont
  {Arbiol}, \citenamefont {i~Morral},\ and\ \citenamefont
  {Grundler}}]{Rueffer2012}%
  \BibitemOpen
  \bibfield  {author} {\bibinfo {author} {\bibfnamefont {D.}~\bibnamefont
  {Rüffer}}, \bibinfo {author} {\bibfnamefont {R.}~\bibnamefont {Huber}},
  \bibinfo {author} {\bibfnamefont {P.}~\bibnamefont {Berberich}}, \bibinfo
  {author} {\bibfnamefont {S.}~\bibnamefont {Albert}}, \bibinfo {author}
  {\bibfnamefont {E.}~\bibnamefont {Russo-Averchi}}, \bibinfo {author}
  {\bibfnamefont {M.}~\bibnamefont {Heiss}}, \bibinfo {author} {\bibfnamefont
  {J.}~\bibnamefont {Arbiol}}, \bibinfo {author} {\bibfnamefont {A.~F.}\
  \bibnamefont {i~Morral}},\ and\ \bibinfo {author} {\bibfnamefont
  {D.}~\bibnamefont {Grundler}},\ }\bibfield  {title} {\bibinfo {title}
  {Magnetic states of an individual \uppercase{N}i nanotube probed by
  anisotropic magnetoresistance},\ }\href {https://doi.org/10.1039/c2nr31086d}
  {\bibfield  {journal} {\bibinfo  {journal} {Nanoscale}\ }\textbf {\bibinfo
  {volume} {4}},\ \bibinfo {pages} {4989} (\bibinfo {year} {2012})}\BibitemShut
  {NoStop}%
\bibitem [{\citenamefont {Wyss}\ \emph {et~al.}(2017)\citenamefont {Wyss},
  \citenamefont {Mehlin}, \citenamefont {Gross}, \citenamefont {Buchter},
  \citenamefont {Farhan}, \citenamefont {Buzzi}, \citenamefont {Kleibert},
  \citenamefont {Tütüncüoglu}, \citenamefont {Heimbach}, \citenamefont
  {i~Morral}, \citenamefont {Grundler},\ and\ \citenamefont
  {Poggio}}]{Wyss2017}%
  \BibitemOpen
  \bibfield  {author} {\bibinfo {author} {\bibfnamefont {M.}~\bibnamefont
  {Wyss}}, \bibinfo {author} {\bibfnamefont {A.}~\bibnamefont {Mehlin}},
  \bibinfo {author} {\bibfnamefont {B.}~\bibnamefont {Gross}}, \bibinfo
  {author} {\bibfnamefont {A.}~\bibnamefont {Buchter}}, \bibinfo {author}
  {\bibfnamefont {A.}~\bibnamefont {Farhan}}, \bibinfo {author} {\bibfnamefont
  {M.}~\bibnamefont {Buzzi}}, \bibinfo {author} {\bibfnamefont
  {A.}~\bibnamefont {Kleibert}}, \bibinfo {author} {\bibfnamefont
  {G.}~\bibnamefont {Tütüncüoglu}}, \bibinfo {author} {\bibfnamefont
  {F.}~\bibnamefont {Heimbach}}, \bibinfo {author} {\bibfnamefont {A.~F.}\
  \bibnamefont {i~Morral}}, \bibinfo {author} {\bibfnamefont {D.}~\bibnamefont
  {Grundler}},\ and\ \bibinfo {author} {\bibfnamefont {M.}~\bibnamefont
  {Poggio}},\ }\bibfield  {title} {\bibinfo {title} {Imaging magnetic vortex
  configurations in ferromagnetic nanotubes},\ }\href
  {https://doi.org/10.1103/physrevb.96.024423} {\bibfield  {journal} {\bibinfo
  {journal} {Phys. Rev. B}\ }\textbf {\bibinfo {volume} {96}},\ \bibinfo
  {pages} {024423} (\bibinfo {year} {2017})}\BibitemShut {NoStop}%
\bibitem [{\citenamefont {Li}\ \emph {et~al.}(2008)\citenamefont {Li},
  \citenamefont {Thompson}, \citenamefont {Bergmann},\ and\ \citenamefont
  {Lu}}]{Li2008}%
  \BibitemOpen
  \bibfield  {author} {\bibinfo {author} {\bibfnamefont {D.}~\bibnamefont
  {Li}}, \bibinfo {author} {\bibfnamefont {R.~S.}\ \bibnamefont {Thompson}},
  \bibinfo {author} {\bibfnamefont {G.}~\bibnamefont {Bergmann}},\ and\
  \bibinfo {author} {\bibfnamefont {J.~G.}\ \bibnamefont {Lu}},\ }\bibfield
  {title} {\bibinfo {title} {Template‐based synthesis and magnetic properties
  of cobalt nanotube arrays},\ }\href {https://doi.org/10.1002/adma.200801455}
  {\bibfield  {journal} {\bibinfo  {journal} {Adv. Mater.}\ }\textbf {\bibinfo
  {volume} {20}},\ \bibinfo {pages} {4575} (\bibinfo {year}
  {2008})}\BibitemShut {NoStop}%
\bibitem [{\citenamefont {Landeros}\ \emph {et~al.}(2007)\citenamefont
  {Landeros}, \citenamefont {Allende}, \citenamefont {Escrig}, \citenamefont
  {Salcedo}, \citenamefont {Altbir},\ and\ \citenamefont
  {Vogel}}]{Landeros2007}%
  \BibitemOpen
  \bibfield  {author} {\bibinfo {author} {\bibfnamefont {P.}~\bibnamefont
  {Landeros}}, \bibinfo {author} {\bibfnamefont {S.}~\bibnamefont {Allende}},
  \bibinfo {author} {\bibfnamefont {J.}~\bibnamefont {Escrig}}, \bibinfo
  {author} {\bibfnamefont {E.}~\bibnamefont {Salcedo}}, \bibinfo {author}
  {\bibfnamefont {D.}~\bibnamefont {Altbir}},\ and\ \bibinfo {author}
  {\bibfnamefont {E.~E.}\ \bibnamefont {Vogel}},\ }\bibfield  {title} {\bibinfo
  {title} {Reversal modes in magnetic nanotubes},\ }\href
  {https://doi.org/10.1063/1.2437655} {\bibfield  {journal} {\bibinfo
  {journal} {Appl. Phys. Lett.}\ }\textbf {\bibinfo {volume} {90}},\ \bibinfo
  {pages} {90} (\bibinfo {year} {2007})}\BibitemShut {NoStop}%
\bibitem [{\citenamefont {Zhang}\ \emph
  {et~al.}(2021{\natexlab{a}})\citenamefont {Zhang}, \citenamefont {Liu},
  \citenamefont {Dong}, \citenamefont {Wu}, \citenamefont {Zhang},
  \citenamefont {Wang}, \citenamefont {Lu}, \citenamefont {Rückriegel},
  \citenamefont {Wang}, \citenamefont {Duine}, \citenamefont {Yu},
  \citenamefont {Luo}, \citenamefont {Shen},\ and\ \citenamefont
  {Zhang}}]{Zhang2021d}%
  \BibitemOpen
  \bibfield  {author} {\bibinfo {author} {\bibfnamefont {Y.}~\bibnamefont
  {Zhang}}, \bibinfo {author} {\bibfnamefont {J.}~\bibnamefont {Liu}}, \bibinfo
  {author} {\bibfnamefont {Y.}~\bibnamefont {Dong}}, \bibinfo {author}
  {\bibfnamefont {S.}~\bibnamefont {Wu}}, \bibinfo {author} {\bibfnamefont
  {J.}~\bibnamefont {Zhang}}, \bibinfo {author} {\bibfnamefont
  {J.}~\bibnamefont {Wang}}, \bibinfo {author} {\bibfnamefont {J.}~\bibnamefont
  {Lu}}, \bibinfo {author} {\bibfnamefont {A.}~\bibnamefont {Rückriegel}},
  \bibinfo {author} {\bibfnamefont {H.}~\bibnamefont {Wang}}, \bibinfo {author}
  {\bibfnamefont {R.}~\bibnamefont {Duine}}, \bibinfo {author} {\bibfnamefont
  {H.}~\bibnamefont {Yu}}, \bibinfo {author} {\bibfnamefont {Z.}~\bibnamefont
  {Luo}}, \bibinfo {author} {\bibfnamefont {K.}~\bibnamefont {Shen}},\ and\
  \bibinfo {author} {\bibfnamefont {J.}~\bibnamefont {Zhang}},\ }\bibfield
  {title} {\bibinfo {title} {Strain-driven dzyaloshinskii-moriya interaction
  for room-temperature magnetic skyrmions},\ }\href
  {https://doi.org/10.1103/physrevlett.127.117204} {\bibfield  {journal}
  {\bibinfo  {journal} {Phys. Rev. Lett.}\ }\textbf {\bibinfo {volume} {127}},\
  \bibinfo {pages} {117204} (\bibinfo {year} {2021}{\natexlab{a}})}\BibitemShut
  {NoStop}%
\bibitem [{\citenamefont {Iijima}(1991)}]{Iijima1991}%
  \BibitemOpen
  \bibfield  {author} {\bibinfo {author} {\bibfnamefont {S.}~\bibnamefont
  {Iijima}},\ }\bibfield  {title} {\bibinfo {title} {Helical microtubules of
  graphitic carbon},\ }\href {https://doi.org/10.1038/354056a0} {\bibfield
  {journal} {\bibinfo  {journal} {Nature}\ }\textbf {\bibinfo {volume} {354}},\
  \bibinfo {pages} {56} (\bibinfo {year} {1991})}\BibitemShut {NoStop}%
\bibitem [{\citenamefont {Wang}\ \emph {et~al.}(2018)\citenamefont {Wang},
  \citenamefont {He},\ and\ \citenamefont {Ding}}]{Wang2018a}%
  \BibitemOpen
  \bibfield  {author} {\bibinfo {author} {\bibfnamefont {X.}~\bibnamefont
  {Wang}}, \bibinfo {author} {\bibfnamefont {M.}~\bibnamefont {He}},\ and\
  \bibinfo {author} {\bibfnamefont {F.}~\bibnamefont {Ding}},\ }\bibfield
  {title} {\bibinfo {title} {Chirality-controlled synthesis of single-walled
  carbon nanotubes—from mechanistic studies toward experimental
  realization},\ }\href {https://doi.org/10.1016/j.mattod.2018.06.001}
  {\bibfield  {journal} {\bibinfo  {journal} {Mater. Today}\ }\textbf {\bibinfo
  {volume} {21}},\ \bibinfo {pages} {845} (\bibinfo {year} {2018})}\BibitemShut
  {NoStop}%
\bibitem [{\citenamefont {Dresselhaus}\ \emph {et~al.}(1996)\citenamefont
  {Dresselhaus}, \citenamefont {Dresselhaus},\ and\ \citenamefont
  {Eklund}}]{CNTbook}%
  \BibitemOpen
  \bibfield  {author} {\bibinfo {author} {\bibfnamefont {M.~S.}\ \bibnamefont
  {Dresselhaus}}, \bibinfo {author} {\bibfnamefont {G.}~\bibnamefont
  {Dresselhaus}},\ and\ \bibinfo {author} {\bibfnamefont {P.~C.}\ \bibnamefont
  {Eklund}},\ }\href@noop {} {\emph {\bibinfo {title} {Science of fullerenes
  and carbon nanotubes: their properties and applications}}}\ (\bibinfo
  {publisher} {Elsevier},\ \bibinfo {year} {1996})\BibitemShut {NoStop}%
\bibitem [{\citenamefont {Han}\ \emph {et~al.}(2009)\citenamefont {Han},
  \citenamefont {Shamaila}, \citenamefont {Sharif}, \citenamefont {Chen},
  \citenamefont {Liu},\ and\ \citenamefont {Liu}}]{Han2009}%
  \BibitemOpen
  \bibfield  {author} {\bibinfo {author} {\bibfnamefont {X.}~\bibnamefont
  {Han}}, \bibinfo {author} {\bibfnamefont {S.}~\bibnamefont {Shamaila}},
  \bibinfo {author} {\bibfnamefont {R.}~\bibnamefont {Sharif}}, \bibinfo
  {author} {\bibfnamefont {J.}~\bibnamefont {Chen}}, \bibinfo {author}
  {\bibfnamefont {H.}~\bibnamefont {Liu}},\ and\ \bibinfo {author}
  {\bibfnamefont {D.}~\bibnamefont {Liu}},\ }\bibfield  {title} {\bibinfo
  {title} {Structural and magnetic properties of various ferromagnetic
  nanotubes},\ }\href {https://doi.org/10.1002/adma.200901065} {\bibfield
  {journal} {\bibinfo  {journal} {Adv. Mater.}\ }\textbf {\bibinfo {volume}
  {21}},\ \bibinfo {pages} {4619} (\bibinfo {year} {2009})}\BibitemShut
  {NoStop}%
\bibitem [{\citenamefont {Korneva}\ \emph {et~al.}(2005)\citenamefont
  {Korneva}, \citenamefont {Ye}, \citenamefont {Gogotsi}, \citenamefont
  {Halverson}, \citenamefont {Friedman}, \citenamefont {Bradley},\ and\
  \citenamefont {Kornev}}]{Korneva2005}%
  \BibitemOpen
  \bibfield  {author} {\bibinfo {author} {\bibfnamefont {G.}~\bibnamefont
  {Korneva}}, \bibinfo {author} {\bibfnamefont {H.}~\bibnamefont {Ye}},
  \bibinfo {author} {\bibfnamefont {Y.}~\bibnamefont {Gogotsi}}, \bibinfo
  {author} {\bibfnamefont {D.}~\bibnamefont {Halverson}}, \bibinfo {author}
  {\bibfnamefont {G.}~\bibnamefont {Friedman}}, \bibinfo {author}
  {\bibfnamefont {J.-C.}\ \bibnamefont {Bradley}},\ and\ \bibinfo {author}
  {\bibfnamefont {K.~G.}\ \bibnamefont {Kornev}},\ }\bibfield  {title}
  {\bibinfo {title} {Carbon nanotubes loaded with magnetic particles},\ }\href
  {https://doi.org/10.1021/nl0502928} {\bibfield  {journal} {\bibinfo
  {journal} {Nano Lett.}\ }\textbf {\bibinfo {volume} {5}},\ \bibinfo {pages}
  {879} (\bibinfo {year} {2005})}\BibitemShut {NoStop}%
\bibitem [{\citenamefont {Lee}\ \emph {et~al.}(2006)\citenamefont {Lee},
  \citenamefont {Cohen},\ and\ \citenamefont {Rubner}}]{Lee2006}%
  \BibitemOpen
  \bibfield  {author} {\bibinfo {author} {\bibfnamefont {D.}~\bibnamefont
  {Lee}}, \bibinfo {author} {\bibfnamefont {R.~E.}\ \bibnamefont {Cohen}},\
  and\ \bibinfo {author} {\bibfnamefont {M.~F.}\ \bibnamefont {Rubner}},\
  }\bibfield  {title} {\bibinfo {title} {Heterostructured magnetic nanotubes},\
  }\href {https://doi.org/10.1021/la0612926} {\bibfield  {journal} {\bibinfo
  {journal} {Langmuir}\ }\textbf {\bibinfo {volume} {23}},\ \bibinfo {pages}
  {123} (\bibinfo {year} {2006})}\BibitemShut {NoStop}%
\bibitem [{\citenamefont {Son}\ \emph {et~al.}(2005)\citenamefont {Son},
  \citenamefont {Reichel}, \citenamefont {He}, \citenamefont {Schuchman},\ and\
  \citenamefont {Lee}}]{Son2005}%
  \BibitemOpen
  \bibfield  {author} {\bibinfo {author} {\bibfnamefont {S.~J.}\ \bibnamefont
  {Son}}, \bibinfo {author} {\bibfnamefont {J.}~\bibnamefont {Reichel}},
  \bibinfo {author} {\bibfnamefont {B.}~\bibnamefont {He}}, \bibinfo {author}
  {\bibfnamefont {M.}~\bibnamefont {Schuchman}},\ and\ \bibinfo {author}
  {\bibfnamefont {S.~B.}\ \bibnamefont {Lee}},\ }\bibfield  {title} {\bibinfo
  {title} {Magnetic nanotubes for magnetic-field-assisted bioseparation,
  biointeraction, and drug delivery},\ }\href
  {https://doi.org/10.1021/ja0517365} {\bibfield  {journal} {\bibinfo
  {journal} {JACS}\ }\textbf {\bibinfo {volume} {127}},\ \bibinfo {pages}
  {7316} (\bibinfo {year} {2005})}\BibitemShut {NoStop}%
\bibitem [{\citenamefont {Guo}\ \emph {et~al.}(2021)\citenamefont {Guo},
  \citenamefont {Jiang}, \citenamefont {Teng}, \citenamefont {Xiong},
  \citenamefont {Chen}, \citenamefont {You},\ and\ \citenamefont
  {Xiao}}]{Guo2021a}%
  \BibitemOpen
  \bibfield  {author} {\bibinfo {author} {\bibfnamefont {J.}~\bibnamefont
  {Guo}}, \bibinfo {author} {\bibfnamefont {H.}~\bibnamefont {Jiang}}, \bibinfo
  {author} {\bibfnamefont {Y.}~\bibnamefont {Teng}}, \bibinfo {author}
  {\bibfnamefont {Y.}~\bibnamefont {Xiong}}, \bibinfo {author} {\bibfnamefont
  {Z.}~\bibnamefont {Chen}}, \bibinfo {author} {\bibfnamefont {L.}~\bibnamefont
  {You}},\ and\ \bibinfo {author} {\bibfnamefont {D.}~\bibnamefont {Xiao}},\
  }\bibfield  {title} {\bibinfo {title} {Recent advances in magnetic carbon
  nanotubes: synthesis, challenges and highlighted applications},\ }\href
  {https://doi.org/10.1039/d1tb01242h} {\bibfield  {journal} {\bibinfo
  {journal} {J. Mater. Chem. B}\ }\textbf {\bibinfo {volume} {9}},\ \bibinfo
  {pages} {9076} (\bibinfo {year} {2021})}\BibitemShut {NoStop}%
\bibitem [{\citenamefont {Giordano}\ \emph {et~al.}(2023)\citenamefont
  {Giordano}, \citenamefont {Hamdi}, \citenamefont {Mucchietto},\ and\
  \citenamefont {Grundler}}]{Giordano2023}%
  \BibitemOpen
  \bibfield  {author} {\bibinfo {author} {\bibfnamefont {M.~C.}\ \bibnamefont
  {Giordano}}, \bibinfo {author} {\bibfnamefont {M.}~\bibnamefont {Hamdi}},
  \bibinfo {author} {\bibfnamefont {A.}~\bibnamefont {Mucchietto}},\ and\
  \bibinfo {author} {\bibfnamefont {D.}~\bibnamefont {Grundler}},\ }\bibfield
  {title} {\bibinfo {title} {Confined spin waves in magnetochiral nanotubes
  with axial and circumferential magnetization},\ }\href
  {https://doi.org/10.1103/physrevmaterials.7.024405} {\bibfield  {journal}
  {\bibinfo  {journal} {Phys. Rev. Materials}\ }\textbf {\bibinfo {volume}
  {7}},\ \bibinfo {pages} {024405} (\bibinfo {year} {2023})}\BibitemShut
  {NoStop}%
\bibitem [{\citenamefont {Shpaisman}\ \emph {et~al.}(2012)\citenamefont
  {Shpaisman}, \citenamefont {Givan}, \citenamefont {Kwiat}, \citenamefont
  {Pevzner}, \citenamefont {Elnathan},\ and\ \citenamefont
  {Patolsky}}]{Shpaisman2012}%
  \BibitemOpen
  \bibfield  {author} {\bibinfo {author} {\bibfnamefont {N.}~\bibnamefont
  {Shpaisman}}, \bibinfo {author} {\bibfnamefont {U.}~\bibnamefont {Givan}},
  \bibinfo {author} {\bibfnamefont {M.}~\bibnamefont {Kwiat}}, \bibinfo
  {author} {\bibfnamefont {A.}~\bibnamefont {Pevzner}}, \bibinfo {author}
  {\bibfnamefont {R.}~\bibnamefont {Elnathan}},\ and\ \bibinfo {author}
  {\bibfnamefont {F.}~\bibnamefont {Patolsky}},\ }\bibfield  {title} {\bibinfo
  {title} {Controlled synthesis of ferromagnetic semiconducting silicon
  nanotubes},\ }\href {https://doi.org/10.1021/jp2037944} {\bibfield  {journal}
  {\bibinfo  {journal} {J. Phys. Chem. C}\ }\textbf {\bibinfo {volume} {116}},\
  \bibinfo {pages} {8000} (\bibinfo {year} {2012})}\BibitemShut {NoStop}%
\bibitem [{\citenamefont {Yan}\ \emph {et~al.}(2012)\citenamefont {Yan},
  \citenamefont {Andreas}, \citenamefont {Kákay}, \citenamefont
  {García-Sánchez},\ and\ \citenamefont {Hertel}}]{Yan2012}%
  \BibitemOpen
  \bibfield  {author} {\bibinfo {author} {\bibfnamefont {M.}~\bibnamefont
  {Yan}}, \bibinfo {author} {\bibfnamefont {C.}~\bibnamefont {Andreas}},
  \bibinfo {author} {\bibfnamefont {A.}~\bibnamefont {Kákay}}, \bibinfo
  {author} {\bibfnamefont {F.}~\bibnamefont {García-Sánchez}},\ and\ \bibinfo
  {author} {\bibfnamefont {R.}~\bibnamefont {Hertel}},\ }\bibfield  {title}
  {\bibinfo {title} {Chiral symmetry breaking and pair-creation mediated walker
  breakdown in magnetic nanotubes},\ }\href {https://doi.org/10.1063/1.4727909}
  {\bibfield  {journal} {\bibinfo  {journal} {Appl. Phys. Lett.}\ }\textbf
  {\bibinfo {volume} {100}},\ \bibinfo {pages} {252401} (\bibinfo {year}
  {2012})}\BibitemShut {NoStop}%
\bibitem [{\citenamefont {Yang}\ \emph {et~al.}(2018)\citenamefont {Yang},
  \citenamefont {Kim}, \citenamefont {Kim}, \citenamefont {Cho}, \citenamefont
  {Lee},\ and\ \citenamefont {Kim}}]{Yang2018}%
  \BibitemOpen
  \bibfield  {author} {\bibinfo {author} {\bibfnamefont {J.}~\bibnamefont
  {Yang}}, \bibinfo {author} {\bibfnamefont {J.}~\bibnamefont {Kim}}, \bibinfo
  {author} {\bibfnamefont {B.}~\bibnamefont {Kim}}, \bibinfo {author}
  {\bibfnamefont {Y.-J.}\ \bibnamefont {Cho}}, \bibinfo {author} {\bibfnamefont
  {J.-H.}\ \bibnamefont {Lee}},\ and\ \bibinfo {author} {\bibfnamefont {S.-K.}\
  \bibnamefont {Kim}},\ }\bibfield  {title} {\bibinfo {title}
  {Vortex-chirality-dependent standing spin-wave modes in soft magnetic
  nanotubes},\ }\href {https://doi.org/10.1063/1.5010405} {\bibfield  {journal}
  {\bibinfo  {journal} {J. Appl. Phys.}\ }\textbf {\bibinfo {volume} {123}},\
  \bibinfo {pages} {033901} (\bibinfo {year} {2018})}\BibitemShut {NoStop}%
\bibitem [{\citenamefont {Körber}\ \emph {et~al.}(2022)\citenamefont
  {Körber}, \citenamefont {Verba}, \citenamefont {Otálora}, \citenamefont
  {Kravchuk}, \citenamefont {Lindner}, \citenamefont {Fassbender},\ and\
  \citenamefont {Kákay}}]{Koerber2022}%
  \BibitemOpen
  \bibfield  {author} {\bibinfo {author} {\bibfnamefont {L.}~\bibnamefont
  {Körber}}, \bibinfo {author} {\bibfnamefont {R.}~\bibnamefont {Verba}},
  \bibinfo {author} {\bibfnamefont {J.~A.}\ \bibnamefont {Otálora}}, \bibinfo
  {author} {\bibfnamefont {V.}~\bibnamefont {Kravchuk}}, \bibinfo {author}
  {\bibfnamefont {J.}~\bibnamefont {Lindner}}, \bibinfo {author} {\bibfnamefont
  {J.}~\bibnamefont {Fassbender}},\ and\ \bibinfo {author} {\bibfnamefont
  {A.}~\bibnamefont {Kákay}},\ }\bibfield  {title} {\bibinfo {title}
  {Curvilinear spin-wave dynamics beyond the thin-shell approximation: Magnetic
  nanotubes as a case study},\ }\href
  {https://doi.org/10.1103/physrevb.106.014405} {\bibfield  {journal} {\bibinfo
   {journal} {Phys. Rev. B}\ }\textbf {\bibinfo {volume} {106}},\ \bibinfo
  {pages} {014405} (\bibinfo {year} {2022})}\BibitemShut {NoStop}%
\bibitem [{\citenamefont {Gallardo}\ \emph {et~al.}(2022)\citenamefont
  {Gallardo}, \citenamefont {Alvarado-Seguel},\ and\ \citenamefont
  {Landeros}}]{Gallardo2022}%
  \BibitemOpen
  \bibfield  {author} {\bibinfo {author} {\bibfnamefont {R.}~\bibnamefont
  {Gallardo}}, \bibinfo {author} {\bibfnamefont {P.}~\bibnamefont
  {Alvarado-Seguel}},\ and\ \bibinfo {author} {\bibfnamefont {P.}~\bibnamefont
  {Landeros}},\ }\bibfield  {title} {\bibinfo {title} {Unidirectional chiral
  magnonics in cylindrical synthetic antiferromagnets},\ }\href
  {https://doi.org/10.1103/physrevapplied.18.054044} {\bibfield  {journal}
  {\bibinfo  {journal} {Phys. Rev. Applied}\ }\textbf {\bibinfo {volume}
  {18}},\ \bibinfo {pages} {054044} (\bibinfo {year} {2022})}\BibitemShut
  {NoStop}%
\bibitem [{\citenamefont {Otálora}\ \emph {et~al.}(2016)\citenamefont
  {Otálora}, \citenamefont {Yan}, \citenamefont {Schultheiss}, \citenamefont
  {Hertel},\ and\ \citenamefont {Kákay}}]{Otalora2016}%
  \BibitemOpen
  \bibfield  {author} {\bibinfo {author} {\bibfnamefont {J.~A.}\ \bibnamefont
  {Otálora}}, \bibinfo {author} {\bibfnamefont {M.}~\bibnamefont {Yan}},
  \bibinfo {author} {\bibfnamefont {H.}~\bibnamefont {Schultheiss}}, \bibinfo
  {author} {\bibfnamefont {R.}~\bibnamefont {Hertel}},\ and\ \bibinfo {author}
  {\bibfnamefont {A.}~\bibnamefont {Kákay}},\ }\bibfield  {title} {\bibinfo
  {title} {Curvature-induced asymmetric spin-wave dispersion},\ }\href
  {https://doi.org/10.1103/physrevlett.117.227203} {\bibfield  {journal}
  {\bibinfo  {journal} {Phys. Rev. Lett.}\ }\textbf {\bibinfo {volume} {117}},\
  \bibinfo {pages} {227203} (\bibinfo {year} {2016})}\BibitemShut {NoStop}%
\bibitem [{\citenamefont {Salazar-Cardona}\ \emph {et~al.}(2021)\citenamefont
  {Salazar-Cardona}, \citenamefont {Körber}, \citenamefont {Schultheiss},
  \citenamefont {Lenz}, \citenamefont {Thomas}, \citenamefont {Nielsch},
  \citenamefont {Kákay},\ and\ \citenamefont {Otálora}}]{SalazarCardona2021}%
  \BibitemOpen
  \bibfield  {author} {\bibinfo {author} {\bibfnamefont {M.~M.}\ \bibnamefont
  {Salazar-Cardona}}, \bibinfo {author} {\bibfnamefont {L.}~\bibnamefont
  {Körber}}, \bibinfo {author} {\bibfnamefont {H.}~\bibnamefont
  {Schultheiss}}, \bibinfo {author} {\bibfnamefont {K.}~\bibnamefont {Lenz}},
  \bibinfo {author} {\bibfnamefont {A.}~\bibnamefont {Thomas}}, \bibinfo
  {author} {\bibfnamefont {K.}~\bibnamefont {Nielsch}}, \bibinfo {author}
  {\bibfnamefont {A.}~\bibnamefont {Kákay}},\ and\ \bibinfo {author}
  {\bibfnamefont {J.~A.}\ \bibnamefont {Otálora}},\ }\bibfield  {title}
  {\bibinfo {title} {Nonreciprocity of spin waves in magnetic nanotubes with
  helical equilibrium magnetization},\ }\href
  {https://doi.org/10.1063/5.0048692} {\bibfield  {journal} {\bibinfo
  {journal} {Appl. Phys. Lett.}\ }\textbf {\bibinfo {volume} {118}},\ \bibinfo
  {pages} {2411} (\bibinfo {year} {2021})}\BibitemShut {NoStop}%
\bibitem [{\citenamefont {Chumak}\ \emph {et~al.}(2015)\citenamefont {Chumak},
  \citenamefont {Vasyuchka}, \citenamefont {Serga},\ and\ \citenamefont
  {Hillebrands}}]{Chumak2015}%
  \BibitemOpen
  \bibfield  {author} {\bibinfo {author} {\bibfnamefont {A.~V.}\ \bibnamefont
  {Chumak}}, \bibinfo {author} {\bibfnamefont {V.}~\bibnamefont {Vasyuchka}},
  \bibinfo {author} {\bibfnamefont {A.}~\bibnamefont {Serga}},\ and\ \bibinfo
  {author} {\bibfnamefont {B.}~\bibnamefont {Hillebrands}},\ }\bibfield
  {title} {\bibinfo {title} {Magnon spintronics},\ }\href
  {https://doi.org/10.1038/nphys3347} {\bibfield  {journal} {\bibinfo
  {journal} {Nat. Phys.}\ }\textbf {\bibinfo {volume} {11}},\ \bibinfo {pages}
  {453} (\bibinfo {year} {2015})}\BibitemShut {NoStop}%
\bibitem [{\citenamefont {Huang}\ \emph {et~al.}(2017)\citenamefont {Huang},
  \citenamefont {Clark}, \citenamefont {Navarro-Moratalla}, \citenamefont
  {Klein}, \citenamefont {Cheng}, \citenamefont {Seyler}, \citenamefont
  {Zhong}, \citenamefont {Schmidgall}, \citenamefont {McGuire}, \citenamefont
  {Cobden}, \citenamefont {Yao}, \citenamefont {Xiao}, \citenamefont
  {Jarillo-Herrero},\ and\ \citenamefont {Xu}}]{Huang2017}%
  \BibitemOpen
  \bibfield  {author} {\bibinfo {author} {\bibfnamefont {B.}~\bibnamefont
  {Huang}}, \bibinfo {author} {\bibfnamefont {G.}~\bibnamefont {Clark}},
  \bibinfo {author} {\bibfnamefont {E.}~\bibnamefont {Navarro-Moratalla}},
  \bibinfo {author} {\bibfnamefont {D.~R.}\ \bibnamefont {Klein}}, \bibinfo
  {author} {\bibfnamefont {R.}~\bibnamefont {Cheng}}, \bibinfo {author}
  {\bibfnamefont {K.~L.}\ \bibnamefont {Seyler}}, \bibinfo {author}
  {\bibfnamefont {D.}~\bibnamefont {Zhong}}, \bibinfo {author} {\bibfnamefont
  {E.}~\bibnamefont {Schmidgall}}, \bibinfo {author} {\bibfnamefont {M.~A.}\
  \bibnamefont {McGuire}}, \bibinfo {author} {\bibfnamefont {D.~H.}\
  \bibnamefont {Cobden}}, \bibinfo {author} {\bibfnamefont {W.}~\bibnamefont
  {Yao}}, \bibinfo {author} {\bibfnamefont {D.}~\bibnamefont {Xiao}}, \bibinfo
  {author} {\bibfnamefont {P.}~\bibnamefont {Jarillo-Herrero}},\ and\ \bibinfo
  {author} {\bibfnamefont {X.}~\bibnamefont {Xu}},\ }\bibfield  {title}
  {\bibinfo {title} {Layer-dependent ferromagnetism in a van der
  \uppercase{W}aals crystal down to the monolayer limit},\ }\href
  {https://doi.org/10.1038/nature22391} {\bibfield  {journal} {\bibinfo
  {journal} {Nature}\ }\textbf {\bibinfo {volume} {546}},\ \bibinfo {pages}
  {270} (\bibinfo {year} {2017})}\BibitemShut {NoStop}%
\bibitem [{\citenamefont {Gong}\ \emph {et~al.}(2017)\citenamefont {Gong},
  \citenamefont {Li}, \citenamefont {Li}, \citenamefont {Ji}, \citenamefont
  {\uppercase{A}lex Stern}, \citenamefont {Xia}, \citenamefont {Cao},
  \citenamefont {Bao}, \citenamefont {Wang}, \citenamefont {Wang},
  \citenamefont {Qiu}, \citenamefont {Cava}, \citenamefont {Louie},
  \citenamefont {Xia},\ and\ \citenamefont {Zhang}}]{Gong2017}%
  \BibitemOpen
  \bibfield  {author} {\bibinfo {author} {\bibfnamefont {C.}~\bibnamefont
  {Gong}}, \bibinfo {author} {\bibfnamefont {L.}~\bibnamefont {Li}}, \bibinfo
  {author} {\bibfnamefont {Z.}~\bibnamefont {Li}}, \bibinfo {author}
  {\bibfnamefont {H.}~\bibnamefont {Ji}}, \bibinfo {author} {\bibnamefont
  {\uppercase{A}lex Stern}}, \bibinfo {author} {\bibfnamefont {Y.}~\bibnamefont
  {Xia}}, \bibinfo {author} {\bibfnamefont {T.}~\bibnamefont {Cao}}, \bibinfo
  {author} {\bibfnamefont {W.}~\bibnamefont {Bao}}, \bibinfo {author}
  {\bibfnamefont {C.}~\bibnamefont {Wang}}, \bibinfo {author} {\bibfnamefont
  {Y.}~\bibnamefont {Wang}}, \bibinfo {author} {\bibfnamefont {Z.~Q.}\
  \bibnamefont {Qiu}}, \bibinfo {author} {\bibfnamefont {R.~J.}\ \bibnamefont
  {Cava}}, \bibinfo {author} {\bibfnamefont {S.~G.}\ \bibnamefont {Louie}},
  \bibinfo {author} {\bibfnamefont {J.}~\bibnamefont {Xia}},\ and\ \bibinfo
  {author} {\bibfnamefont {X.}~\bibnamefont {Zhang}},\ }\bibfield  {title}
  {\bibinfo {title} {Discovery of intrinsic ferromagnetism in two-dimensional
  van der \uppercase{W}aals crystals},\ }\href
  {https://doi.org/10.1038/nature22060} {\bibfield  {journal} {\bibinfo
  {journal} {Nature}\ }\textbf {\bibinfo {volume} {546}},\ \bibinfo {pages}
  {265} (\bibinfo {year} {2017})}\BibitemShut {NoStop}%
\bibitem [{\citenamefont {Achinuq}\ \emph {et~al.}(2021)\citenamefont
  {Achinuq}, \citenamefont {Fujita}, \citenamefont {Xia}, \citenamefont {Guo},
  \citenamefont {Bencok}, \citenamefont {van~der Laan},\ and\ \citenamefont
  {Hesjedal}}]{Achinuq2021}%
  \BibitemOpen
  \bibfield  {author} {\bibinfo {author} {\bibfnamefont {B.}~\bibnamefont
  {Achinuq}}, \bibinfo {author} {\bibfnamefont {R.}~\bibnamefont {Fujita}},
  \bibinfo {author} {\bibfnamefont {W.}~\bibnamefont {Xia}}, \bibinfo {author}
  {\bibfnamefont {Y.}~\bibnamefont {Guo}}, \bibinfo {author} {\bibfnamefont
  {P.}~\bibnamefont {Bencok}}, \bibinfo {author} {\bibfnamefont
  {G.}~\bibnamefont {van~der Laan}},\ and\ \bibinfo {author} {\bibfnamefont
  {T.}~\bibnamefont {Hesjedal}},\ }\bibfield  {title} {\bibinfo {title}
  {Covalent mixing in the 2\uppercase{D} ferromagnet
  \uppercase{C}r\uppercase{S}i\uppercase{T}e$_3$ evidenced by magnetic
  \uppercase{X}-ray circular dichroism},\ }\href
  {https://doi.org/10.1002/pssr.202100566} {\bibfield  {journal} {\bibinfo
  {journal} {Phys. Status. Solidi.}\ }\textbf {\bibinfo {volume} {16}},\
  \bibinfo {pages} {2100566} (\bibinfo {year} {2021})}\BibitemShut {NoStop}%
\bibitem [{\citenamefont {Lee}\ \emph {et~al.}(2021)\citenamefont {Lee},
  \citenamefont {Dismukes}, \citenamefont {Telford}, \citenamefont {Wiscons},
  \citenamefont {Wang}, \citenamefont {Xu}, \citenamefont {Nuckolls},
  \citenamefont {Dean}, \citenamefont {Roy},\ and\ \citenamefont
  {Zhu}}]{Lee2021}%
  \BibitemOpen
  \bibfield  {author} {\bibinfo {author} {\bibfnamefont {K.}~\bibnamefont
  {Lee}}, \bibinfo {author} {\bibfnamefont {A.~H.}\ \bibnamefont {Dismukes}},
  \bibinfo {author} {\bibfnamefont {E.~J.}\ \bibnamefont {Telford}}, \bibinfo
  {author} {\bibfnamefont {R.~A.}\ \bibnamefont {Wiscons}}, \bibinfo {author}
  {\bibfnamefont {J.}~\bibnamefont {Wang}}, \bibinfo {author} {\bibfnamefont
  {X.}~\bibnamefont {Xu}}, \bibinfo {author} {\bibfnamefont {C.}~\bibnamefont
  {Nuckolls}}, \bibinfo {author} {\bibfnamefont {C.~R.}\ \bibnamefont {Dean}},
  \bibinfo {author} {\bibfnamefont {X.}~\bibnamefont {Roy}},\ and\ \bibinfo
  {author} {\bibfnamefont {X.}~\bibnamefont {Zhu}},\ }\bibfield  {title}
  {\bibinfo {title} {Magnetic order and symmetry in the 2\uppercase{D}
  semiconductor \uppercase{C}r\uppercase{SB}r},\ }\href
  {https://doi.org/10.1021/acs.nanolett.1c00219} {\bibfield  {journal}
  {\bibinfo  {journal} {Nano Lett.}\ }\textbf {\bibinfo {volume} {21}},\
  \bibinfo {pages} {3511} (\bibinfo {year} {2021})}\BibitemShut {NoStop}%
\bibitem [{\citenamefont {Zhang}\ \emph
  {et~al.}(2021{\natexlab{b}})\citenamefont {Zhang}, \citenamefont {Lu},
  \citenamefont {Liu}, \citenamefont {\uppercase{N}iu}, \citenamefont {Sun},
  \citenamefont {\uppercase{C}o\uppercase{o}k}, \citenamefont {Vaninger},
  \citenamefont {Miceli}, \citenamefont {Singh}, \citenamefont {Lian},
  \citenamefont {Chang}, \citenamefont {He}, \citenamefont {Du}, \citenamefont
  {He}, \citenamefont {Zhang}, \citenamefont {Bian},\ and\ \citenamefont
  {Xu}}]{Zhang2021}%
  \BibitemOpen
  \bibfield  {author} {\bibinfo {author} {\bibfnamefont {X.}~\bibnamefont
  {Zhang}}, \bibinfo {author} {\bibfnamefont {Q.}~\bibnamefont {Lu}}, \bibinfo
  {author} {\bibfnamefont {W.}~\bibnamefont {Liu}}, \bibinfo {author}
  {\bibfnamefont {W.}~\bibnamefont {\uppercase{N}iu}}, \bibinfo {author}
  {\bibfnamefont {J.}~\bibnamefont {Sun}}, \bibinfo {author} {\bibfnamefont
  {J.}~\bibnamefont {\uppercase{C}o\uppercase{o}k}}, \bibinfo {author}
  {\bibfnamefont {M.}~\bibnamefont {Vaninger}}, \bibinfo {author}
  {\bibfnamefont {P.~F.}\ \bibnamefont {Miceli}}, \bibinfo {author}
  {\bibfnamefont {D.~J.}\ \bibnamefont {Singh}}, \bibinfo {author}
  {\bibfnamefont {S.-W.}\ \bibnamefont {Lian}}, \bibinfo {author}
  {\bibfnamefont {T.-R.}\ \bibnamefont {Chang}}, \bibinfo {author}
  {\bibfnamefont {X.}~\bibnamefont {He}}, \bibinfo {author} {\bibfnamefont
  {J.}~\bibnamefont {Du}}, \bibinfo {author} {\bibfnamefont {L.}~\bibnamefont
  {He}}, \bibinfo {author} {\bibfnamefont {R.}~\bibnamefont {Zhang}}, \bibinfo
  {author} {\bibfnamefont {G.}~\bibnamefont {Bian}},\ and\ \bibinfo {author}
  {\bibfnamefont {Y.}~\bibnamefont {Xu}},\ }\bibfield  {title} {\bibinfo
  {title} {Room-temperature intrinsic ferromagnetism in epitaxial
  \uppercase{C}r\uppercase{T}e$_2$ ultrathin films},\ }\href
  {https://doi.org/10.1038/s41467-021-22777-x} {\bibfield  {journal} {\bibinfo
  {journal} {Nat. Commun.}\ }\textbf {\bibinfo {volume} {12}},\ \bibinfo
  {pages} {2492} (\bibinfo {year} {2021}{\natexlab{b}})}\BibitemShut {NoStop}%
\bibitem [{\citenamefont {Xian}\ \emph {et~al.}(2022)\citenamefont {Xian},
  \citenamefont {Wang}, \citenamefont {Nie}, \citenamefont {Li}, \citenamefont
  {Han}, \citenamefont {Lin}, \citenamefont {Zhang}, \citenamefont {Liu},
  \citenamefont {Zhang}, \citenamefont {Miao}, \citenamefont {Yi},
  \citenamefont {Wu}, \citenamefont {Chen}, \citenamefont {Han}, \citenamefont
  {Xia}, \citenamefont {Ji},\ and\ \citenamefont {Fu}}]{Xian2022}%
  \BibitemOpen
  \bibfield  {author} {\bibinfo {author} {\bibfnamefont {J.-J.}\ \bibnamefont
  {Xian}}, \bibinfo {author} {\bibfnamefont {C.}~\bibnamefont {Wang}}, \bibinfo
  {author} {\bibfnamefont {J.-H.}\ \bibnamefont {Nie}}, \bibinfo {author}
  {\bibfnamefont {R.}~\bibnamefont {Li}}, \bibinfo {author} {\bibfnamefont
  {M.}~\bibnamefont {Han}}, \bibinfo {author} {\bibfnamefont {J.}~\bibnamefont
  {Lin}}, \bibinfo {author} {\bibfnamefont {W.-H.}\ \bibnamefont {Zhang}},
  \bibinfo {author} {\bibfnamefont {Z.-Y.}\ \bibnamefont {Liu}}, \bibinfo
  {author} {\bibfnamefont {Z.-M.}\ \bibnamefont {Zhang}}, \bibinfo {author}
  {\bibfnamefont {M.-P.}\ \bibnamefont {Miao}}, \bibinfo {author}
  {\bibfnamefont {Y.}~\bibnamefont {Yi}}, \bibinfo {author} {\bibfnamefont
  {S.}~\bibnamefont {Wu}}, \bibinfo {author} {\bibfnamefont {X.}~\bibnamefont
  {Chen}}, \bibinfo {author} {\bibfnamefont {J.}~\bibnamefont {Han}}, \bibinfo
  {author} {\bibfnamefont {Z.}~\bibnamefont {Xia}}, \bibinfo {author}
  {\bibfnamefont {W.}~\bibnamefont {Ji}},\ and\ \bibinfo {author}
  {\bibfnamefont {Y.-S.}\ \bibnamefont {Fu}},\ }\bibfield  {title} {\bibinfo
  {title} {Spin mapping of intralayer antiferromagnetism and field-induced spin
  reorientation in monolayer \uppercase{C}r\uppercase{T}e$_2$},\ }\href
  {https://doi.org/10.1038/s41467-021-27834-z} {\bibfield  {journal} {\bibinfo
  {journal} {Nat. Commun.}\ }\textbf {\bibinfo {volume} {13}},\ \bibinfo
  {pages} {13:257} (\bibinfo {year} {2022})}\BibitemShut {NoStop}%
\bibitem [{\citenamefont {Wang}\ \emph {et~al.}(2024)\citenamefont {Wang},
  \citenamefont {Wang}, \citenamefont {Hu}, \citenamefont {Wang}, \citenamefont
  {Zou}, \citenamefont {Guo}, \citenamefont {Li}, \citenamefont {Wang},
  \citenamefont {Li}, \citenamefont {Song}, \citenamefont {Wang},\ and\
  \citenamefont {Liu}}]{Wang2024}%
  \BibitemOpen
  \bibfield  {author} {\bibinfo {author} {\bibfnamefont {D.}~\bibnamefont
  {Wang}}, \bibinfo {author} {\bibfnamefont {X.}~\bibnamefont {Wang}}, \bibinfo
  {author} {\bibfnamefont {B.}~\bibnamefont {Hu}}, \bibinfo {author}
  {\bibfnamefont {J.}~\bibnamefont {Wang}}, \bibinfo {author} {\bibfnamefont
  {Y.}~\bibnamefont {Zou}}, \bibinfo {author} {\bibfnamefont {J.}~\bibnamefont
  {Guo}}, \bibinfo {author} {\bibfnamefont {Z.}~\bibnamefont {Li}}, \bibinfo
  {author} {\bibfnamefont {S.}~\bibnamefont {Wang}}, \bibinfo {author}
  {\bibfnamefont {Y.}~\bibnamefont {Li}}, \bibinfo {author} {\bibfnamefont
  {G.}~\bibnamefont {Song}}, \bibinfo {author} {\bibfnamefont {H.}~\bibnamefont
  {Wang}},\ and\ \bibinfo {author} {\bibfnamefont {Y.}~\bibnamefont {Liu}},\
  }\bibfield  {title} {\bibinfo {title} {Strain- and electron doping-induced
  in-plane spin orientation at room temperature in single-layer
  \uppercase{C}r\uppercase{T}e$_2$},\ }\href
  {https://doi.org/10.1021/acsami.4c01034} {\bibfield  {journal} {\bibinfo
  {journal} {ACS Appl. Mater. Interfaces}\ ,\ \bibinfo {pages} {28791}}
  (\bibinfo {year} {2024})}\BibitemShut {NoStop}%
\bibitem [{\citenamefont {Chua}\ \emph {et~al.}(2021)\citenamefont {Chua},
  \citenamefont {Zhou}, \citenamefont {Yu}, \citenamefont {Yu}, \citenamefont
  {Gou}, \citenamefont {Zhu}, \citenamefont {Zhang}, \citenamefont {Liu},
  \citenamefont {Breese}, \citenamefont {Chen}, \citenamefont {Loh},
  \citenamefont {Feng}, \citenamefont {Yang}, \citenamefont {Huang},\ and\
  \citenamefont {Wee}}]{Chua2021}%
  \BibitemOpen
  \bibfield  {author} {\bibinfo {author} {\bibfnamefont {R.}~\bibnamefont
  {Chua}}, \bibinfo {author} {\bibfnamefont {J.}~\bibnamefont {Zhou}}, \bibinfo
  {author} {\bibfnamefont {X.}~\bibnamefont {Yu}}, \bibinfo {author}
  {\bibfnamefont {W.}~\bibnamefont {Yu}}, \bibinfo {author} {\bibfnamefont
  {J.}~\bibnamefont {Gou}}, \bibinfo {author} {\bibfnamefont {R.}~\bibnamefont
  {Zhu}}, \bibinfo {author} {\bibfnamefont {L.}~\bibnamefont {Zhang}}, \bibinfo
  {author} {\bibfnamefont {M.}~\bibnamefont {Liu}}, \bibinfo {author}
  {\bibfnamefont {M.~B.~H.}\ \bibnamefont {Breese}}, \bibinfo {author}
  {\bibfnamefont {W.}~\bibnamefont {Chen}}, \bibinfo {author} {\bibfnamefont
  {K.~P.}\ \bibnamefont {Loh}}, \bibinfo {author} {\bibfnamefont {Y.~P.}\
  \bibnamefont {Feng}}, \bibinfo {author} {\bibfnamefont {M.}~\bibnamefont
  {Yang}}, \bibinfo {author} {\bibfnamefont {Y.~L.}\ \bibnamefont {Huang}},\
  and\ \bibinfo {author} {\bibfnamefont {A.~T.~S.}\ \bibnamefont {Wee}},\
  }\bibfield  {title} {\bibinfo {title} {Room temperature ferromagnetism of
  monolayer chromium telluride with perpendicular magnetic anisotropy},\ }\href
  {https://doi.org/10.1002/adma.202103360} {\bibfield  {journal} {\bibinfo
  {journal} {Adv. Mater.}\ }\textbf {\bibinfo {volume} {33}},\ \bibinfo {pages}
  {2103360} (\bibinfo {year} {2021})}\BibitemShut {NoStop}%
\bibitem [{\citenamefont {Li}\ \emph {et~al.}(2022{\natexlab{a}})\citenamefont
  {Li}, \citenamefont {Deng}, \citenamefont {Shu}, \citenamefont {Cheng},
  \citenamefont {Qian}, \citenamefont {Wan}, \citenamefont {Zhao},
  \citenamefont {Shen}, \citenamefont {Wu}, \citenamefont {Shi}, \citenamefont
  {Zhang}, \citenamefont {Zhang}, \citenamefont {Yang}, \citenamefont {Zhang},
  \citenamefont {Zhong}, \citenamefont {Xia}, \citenamefont {Li}, \citenamefont
  {Liu}, \citenamefont {Liao}, \citenamefont {Ye}, \citenamefont {Dai},
  \citenamefont {Peng}, \citenamefont {Li},\ and\ \citenamefont
  {Duan}}]{Li2022b}%
  \BibitemOpen
  \bibfield  {author} {\bibinfo {author} {\bibfnamefont {B.}~\bibnamefont
  {Li}}, \bibinfo {author} {\bibfnamefont {X.}~\bibnamefont {Deng}}, \bibinfo
  {author} {\bibfnamefont {W.}~\bibnamefont {Shu}}, \bibinfo {author}
  {\bibfnamefont {X.}~\bibnamefont {Cheng}}, \bibinfo {author} {\bibfnamefont
  {Q.}~\bibnamefont {Qian}}, \bibinfo {author} {\bibfnamefont {Z.}~\bibnamefont
  {Wan}}, \bibinfo {author} {\bibfnamefont {B.}~\bibnamefont {Zhao}}, \bibinfo
  {author} {\bibfnamefont {X.}~\bibnamefont {Shen}}, \bibinfo {author}
  {\bibfnamefont {R.}~\bibnamefont {Wu}}, \bibinfo {author} {\bibfnamefont
  {S.}~\bibnamefont {Shi}}, \bibinfo {author} {\bibfnamefont {H.}~\bibnamefont
  {Zhang}}, \bibinfo {author} {\bibfnamefont {Z.}~\bibnamefont {Zhang}},
  \bibinfo {author} {\bibfnamefont {X.}~\bibnamefont {Yang}}, \bibinfo {author}
  {\bibfnamefont {J.}~\bibnamefont {Zhang}}, \bibinfo {author} {\bibfnamefont
  {M.}~\bibnamefont {Zhong}}, \bibinfo {author} {\bibfnamefont
  {Q.}~\bibnamefont {Xia}}, \bibinfo {author} {\bibfnamefont {J.}~\bibnamefont
  {Li}}, \bibinfo {author} {\bibfnamefont {Y.}~\bibnamefont {Liu}}, \bibinfo
  {author} {\bibfnamefont {L.}~\bibnamefont {Liao}}, \bibinfo {author}
  {\bibfnamefont {Y.}~\bibnamefont {Ye}}, \bibinfo {author} {\bibfnamefont
  {L.}~\bibnamefont {Dai}}, \bibinfo {author} {\bibfnamefont {Y.}~\bibnamefont
  {Peng}}, \bibinfo {author} {\bibfnamefont {B.}~\bibnamefont {Li}},\ and\
  \bibinfo {author} {\bibfnamefont {X.}~\bibnamefont {Duan}},\ }\bibfield
  {title} {\bibinfo {title} {Air-stable ultrathin
  \uppercase{C}r$_3$\uppercase{T}e$_4$ nanosheets with thickness-dependent
  magnetic biskyrmions},\ }\href {https://doi.org/10.1016/j.mattod.2022.04.011}
  {\bibfield  {journal} {\bibinfo  {journal} {Mater. Today}\ }\textbf {\bibinfo
  {volume} {57}},\ \bibinfo {pages} {66} (\bibinfo {year}
  {2022}{\natexlab{a}})}\BibitemShut {NoStop}%
\bibitem [{\citenamefont {Seo}\ \emph {et~al.}(2020)\citenamefont {Seo},
  \citenamefont {Kim}, \citenamefont {An}, \citenamefont {Kim}, \citenamefont
  {Kim}, \citenamefont {Hwang}, \citenamefont {Kim}, \citenamefont {Jang},
  \citenamefont {Kim}, \citenamefont {Eom}, \citenamefont {Seo}, \citenamefont
  {Stania}, \citenamefont {Muntwiler}, \citenamefont {Lee}, \citenamefont
  {Watanabe}, \citenamefont {Taniguchi}, \citenamefont {Jo}, \citenamefont
  {Lee}, \citenamefont {Min}, \citenamefont {Jo}, \citenamefont {Yeom},
  \citenamefont {Choi}, \citenamefont {Shim},\ and\ \citenamefont
  {Kim}}]{Seo2020}%
  \BibitemOpen
  \bibfield  {author} {\bibinfo {author} {\bibfnamefont {J.}~\bibnamefont
  {Seo}}, \bibinfo {author} {\bibfnamefont {D.}~\bibnamefont {Kim}}, \bibinfo
  {author} {\bibfnamefont {E.}~\bibnamefont {An}}, \bibinfo {author}
  {\bibfnamefont {K.}~\bibnamefont {Kim}}, \bibinfo {author} {\bibfnamefont
  {G.-Y.}\ \bibnamefont {Kim}}, \bibinfo {author} {\bibfnamefont {S.-Y.}\
  \bibnamefont {Hwang}}, \bibinfo {author} {\bibfnamefont {D.}~\bibnamefont
  {Kim}}, \bibinfo {author} {\bibfnamefont {B.}~\bibnamefont {Jang}}, \bibinfo
  {author} {\bibfnamefont {H.}~\bibnamefont {Kim}}, \bibinfo {author}
  {\bibfnamefont {G.}~\bibnamefont {Eom}}, \bibinfo {author} {\bibfnamefont
  {S.}~\bibnamefont {Seo}}, \bibinfo {author} {\bibfnamefont {R.}~\bibnamefont
  {Stania}}, \bibinfo {author} {\bibfnamefont {M.}~\bibnamefont {Muntwiler}},
  \bibinfo {author} {\bibfnamefont {J.}~\bibnamefont {Lee}}, \bibinfo {author}
  {\bibfnamefont {K.}~\bibnamefont {Watanabe}}, \bibinfo {author}
  {\bibfnamefont {T.}~\bibnamefont {Taniguchi}}, \bibinfo {author}
  {\bibfnamefont {Y.}~\bibnamefont {Jo}}, \bibinfo {author} {\bibfnamefont
  {J.}~\bibnamefont {Lee}}, \bibinfo {author} {\bibfnamefont {B.}~\bibnamefont
  {Min}}, \bibinfo {author} {\bibfnamefont {M.}~\bibnamefont {Jo}}, \bibinfo
  {author} {\bibfnamefont {H.}~\bibnamefont {Yeom}}, \bibinfo {author}
  {\bibfnamefont {S.-Y.}\ \bibnamefont {Choi}}, \bibinfo {author}
  {\bibfnamefont {J.}~\bibnamefont {Shim}},\ and\ \bibinfo {author}
  {\bibfnamefont {J.}~\bibnamefont {Kim}},\ }\bibfield  {title} {\bibinfo
  {title} {Nearly room temperature ferromagnetism in a magnetic metal-rich van
  der \uppercase{W}aals metal},\ }\href
  {https://doi.org/10.1126/sciadv.aay8912} {\bibfield  {journal} {\bibinfo
  {journal} {Sci. Adv.}\ }\textbf {\bibinfo {volume} {6}},\ \bibinfo {pages}
  {eaay8912} (\bibinfo {year} {2020})}\BibitemShut {NoStop}%
\bibitem [{\citenamefont {Zhang}\ \emph {et~al.}(2022)\citenamefont {Zhang},
  \citenamefont {Guo}, \citenamefont {Wu}, \citenamefont {Wen}, \citenamefont
  {Yang}, \citenamefont {Jin}, \citenamefont {Zhang},\ and\ \citenamefont
  {Chang}}]{Zhang2022}%
  \BibitemOpen
  \bibfield  {author} {\bibinfo {author} {\bibfnamefont {G.}~\bibnamefont
  {Zhang}}, \bibinfo {author} {\bibfnamefont {F.}~\bibnamefont {Guo}}, \bibinfo
  {author} {\bibfnamefont {H.}~\bibnamefont {Wu}}, \bibinfo {author}
  {\bibfnamefont {X.}~\bibnamefont {Wen}}, \bibinfo {author} {\bibfnamefont
  {L.}~\bibnamefont {Yang}}, \bibinfo {author} {\bibfnamefont {W.}~\bibnamefont
  {Jin}}, \bibinfo {author} {\bibfnamefont {W.}~\bibnamefont {Zhang}},\ and\
  \bibinfo {author} {\bibfnamefont {H.}~\bibnamefont {Chang}},\ }\bibfield
  {title} {\bibinfo {title} {Above-room-temperature strong intrinsic
  ferromagnetism in 2\uppercase{D} van der \uppercase{W}aals
  \uppercase{F}e$_3$\uppercase{G}a\uppercase{T}e$_2$ with large perpendicular
  magnetic anisotropy},\ }\href {https://doi.org/10.1038/s41467-022-32605-5}
  {\bibfield  {journal} {\bibinfo  {journal} {Nat. Commun.}\ }\textbf {\bibinfo
  {volume} {13}},\ \bibinfo {pages} {5067} (\bibinfo {year}
  {2022})}\BibitemShut {NoStop}%
\bibitem [{\citenamefont {Chen}\ \emph {et~al.}(2024)\citenamefont {Chen},
  \citenamefont {Yang}, \citenamefont {Ying},\ and\ \citenamefont
  {Guo}}]{Chen2024}%
  \BibitemOpen
  \bibfield  {author} {\bibinfo {author} {\bibfnamefont {Z.}~\bibnamefont
  {Chen}}, \bibinfo {author} {\bibfnamefont {Y.}~\bibnamefont {Yang}}, \bibinfo
  {author} {\bibfnamefont {T.}~\bibnamefont {Ying}},\ and\ \bibinfo {author}
  {\bibfnamefont {J.-g.}\ \bibnamefont {Guo}},\ }\bibfield  {title} {\bibinfo
  {title} {High-\uppercase{T}$_c$ ferromagnetic semiconductor in thinned
  3\uppercase{D} \uppercase{I}sing ferromagnetic metal
  \uppercase{F}e$_3$\uppercase{G}a\uppercase{T}e$_2$},\ }\href
  {https://doi.org/10.1021/acs.nanolett.3c04462} {\bibfield  {journal}
  {\bibinfo  {journal} {Nano Lett.}\ }\textbf {\bibinfo {volume} {24}},\
  \bibinfo {pages} {993} (\bibinfo {year} {2024})}\BibitemShut {NoStop}%
\bibitem [{\citenamefont {Xiao}\ \emph {et~al.}(2022)\citenamefont {Xiao},
  \citenamefont {Zhuang}, \citenamefont {Loh}, \citenamefont {Liang},
  \citenamefont {Gayen}, \citenamefont {Ye}, \citenamefont {Bosman},
  \citenamefont {Eda}, \citenamefont {Wang},\ and\ \citenamefont
  {Xu}}]{Xiao2022}%
  \BibitemOpen
  \bibfield  {author} {\bibinfo {author} {\bibfnamefont {H.}~\bibnamefont
  {Xiao}}, \bibinfo {author} {\bibfnamefont {W.}~\bibnamefont {Zhuang}},
  \bibinfo {author} {\bibfnamefont {L.}~\bibnamefont {Loh}}, \bibinfo {author}
  {\bibfnamefont {T.}~\bibnamefont {Liang}}, \bibinfo {author} {\bibfnamefont
  {A.}~\bibnamefont {Gayen}}, \bibinfo {author} {\bibfnamefont
  {P.}~\bibnamefont {Ye}}, \bibinfo {author} {\bibfnamefont {M.}~\bibnamefont
  {Bosman}}, \bibinfo {author} {\bibfnamefont {G.}~\bibnamefont {Eda}},
  \bibinfo {author} {\bibfnamefont {X.}~\bibnamefont {Wang}},\ and\ \bibinfo
  {author} {\bibfnamefont {M.}~\bibnamefont {Xu}},\ }\bibfield  {title}
  {\bibinfo {title} {Van der \uppercase{W}aals epitaxial growth of
  2\uppercase{D} layered room‐temperature ferromagnetic
  \uppercase{C}r\uppercase{S}$_2$},\ }\href
  {https://doi.org/10.1002/admi.202201353} {\bibfield  {journal} {\bibinfo
  {journal} {Adv. Mater. Interfaces}\ }\textbf {\bibinfo {volume} {9}},\
  \bibinfo {pages} {2201353} (\bibinfo {year} {2022})}\BibitemShut {NoStop}%
\bibitem [{\citenamefont {Bonilla}\ \emph {et~al.}(2018)\citenamefont
  {Bonilla}, \citenamefont {Kolekar}, \citenamefont {Ma}, \citenamefont {Diaz},
  \citenamefont {Kalappattil}, \citenamefont {Das}, \citenamefont {Eggers},
  \citenamefont {Gutierrez}, \citenamefont {Phan},\ and\ \citenamefont
  {Batzill}}]{Bonilla2018}%
  \BibitemOpen
  \bibfield  {author} {\bibinfo {author} {\bibfnamefont {M.}~\bibnamefont
  {Bonilla}}, \bibinfo {author} {\bibfnamefont {S.}~\bibnamefont {Kolekar}},
  \bibinfo {author} {\bibfnamefont {Y.}~\bibnamefont {Ma}}, \bibinfo {author}
  {\bibfnamefont {H.~C.}\ \bibnamefont {Diaz}}, \bibinfo {author}
  {\bibfnamefont {V.}~\bibnamefont {Kalappattil}}, \bibinfo {author}
  {\bibfnamefont {R.}~\bibnamefont {Das}}, \bibinfo {author} {\bibfnamefont
  {T.}~\bibnamefont {Eggers}}, \bibinfo {author} {\bibfnamefont {H.~R.}\
  \bibnamefont {Gutierrez}}, \bibinfo {author} {\bibfnamefont {M.-H.}\
  \bibnamefont {Phan}},\ and\ \bibinfo {author} {\bibfnamefont
  {M.}~\bibnamefont {Batzill}},\ }\bibfield  {title} {\bibinfo {title} {Strong
  room-temperature ferromagnetism in \uppercase{VS}e$_2$ monolayers on van der
  \uppercase{W}aals substrates},\ }\href
  {https://doi.org/10.1038/s41565-018-0063-9} {\bibfield  {journal} {\bibinfo
  {journal} {Nat. Nanotechnol.}\ }\textbf {\bibinfo {volume} {13}},\ \bibinfo
  {pages} {289} (\bibinfo {year} {2018})}\BibitemShut {NoStop}%
\bibitem [{\citenamefont {Meng}\ \emph {et~al.}(2021)\citenamefont {Meng},
  \citenamefont {Zhou}, \citenamefont {Xu}, \citenamefont {Yang}, \citenamefont
  {Si}, \citenamefont {Liu}, \citenamefont {Wang}, \citenamefont {Jiang},
  \citenamefont {Li}, \citenamefont {Qin}, \citenamefont {Zhang}, \citenamefont
  {Wang}, \citenamefont {Liu}, \citenamefont {Tang}, \citenamefont {Ye},
  \citenamefont {Zhou}, \citenamefont {Bao}, \citenamefont {Gao},\ and\
  \citenamefont {Gong}}]{Meng2021}%
  \BibitemOpen
  \bibfield  {author} {\bibinfo {author} {\bibfnamefont {L.}~\bibnamefont
  {Meng}}, \bibinfo {author} {\bibfnamefont {Z.}~\bibnamefont {Zhou}}, \bibinfo
  {author} {\bibfnamefont {M.}~\bibnamefont {Xu}}, \bibinfo {author}
  {\bibfnamefont {S.}~\bibnamefont {Yang}}, \bibinfo {author} {\bibfnamefont
  {K.}~\bibnamefont {Si}}, \bibinfo {author} {\bibfnamefont {L.}~\bibnamefont
  {Liu}}, \bibinfo {author} {\bibfnamefont {X.}~\bibnamefont {Wang}}, \bibinfo
  {author} {\bibfnamefont {H.}~\bibnamefont {Jiang}}, \bibinfo {author}
  {\bibfnamefont {B.}~\bibnamefont {Li}}, \bibinfo {author} {\bibfnamefont
  {P.}~\bibnamefont {Qin}}, \bibinfo {author} {\bibfnamefont {P.}~\bibnamefont
  {Zhang}}, \bibinfo {author} {\bibfnamefont {J.}~\bibnamefont {Wang}},
  \bibinfo {author} {\bibfnamefont {Z.}~\bibnamefont {Liu}}, \bibinfo {author}
  {\bibfnamefont {P.}~\bibnamefont {Tang}}, \bibinfo {author} {\bibfnamefont
  {Y.}~\bibnamefont {Ye}}, \bibinfo {author} {\bibfnamefont {W.}~\bibnamefont
  {Zhou}}, \bibinfo {author} {\bibfnamefont {L.}~\bibnamefont {Bao}}, \bibinfo
  {author} {\bibfnamefont {H.-J.}\ \bibnamefont {Gao}},\ and\ \bibinfo {author}
  {\bibfnamefont {Y.}~\bibnamefont {Gong}},\ }\bibfield  {title} {\bibinfo
  {title} {Anomalous thickness dependence of \uppercase{C}urie temperature in
  air-stable two-dimensional ferromagnetic
  1\uppercase{T}-\uppercase{C}r\uppercase{T}e$_2$ grown by chemical vapor
  deposition},\ }\href {https://doi.org/10.1038/s41467-021-21072-z} {\bibfield
  {journal} {\bibinfo  {journal} {Nat. Commun.}\ }\textbf {\bibinfo {volume}
  {12}},\ \bibinfo {pages} {94} (\bibinfo {year} {2021})}\BibitemShut {NoStop}%
\bibitem [{\citenamefont {Aftab}\ \emph {et~al.}(2022)\citenamefont {Aftab},
  \citenamefont {Iqbal},\ and\ \citenamefont {Rim}}]{Aftab2022}%
  \BibitemOpen
  \bibfield  {author} {\bibinfo {author} {\bibfnamefont {S.}~\bibnamefont
  {Aftab}}, \bibinfo {author} {\bibfnamefont {M.~Z.}\ \bibnamefont {Iqbal}},\
  and\ \bibinfo {author} {\bibfnamefont {Y.~S.}\ \bibnamefont {Rim}},\
  }\bibfield  {title} {\bibinfo {title} {Recent advances in rolling
  2\uppercase{D} \uppercase{TMD}s nanosheets into 1\uppercase{D}
  \uppercase{TMD}s nanotubes/nanoscrolls},\ }\href
  {https://doi.org/10.1002/smll.202205418} {\bibfield  {journal} {\bibinfo
  {journal} {Small}\ }\textbf {\bibinfo {volume} {19}},\ \bibinfo {pages}
  {2205418} (\bibinfo {year} {2022})}\BibitemShut {NoStop}%
\bibitem [{\citenamefont {Shubina}\ \emph {et~al.}(2019)\citenamefont
  {Shubina}, \citenamefont {Remškar}, \citenamefont {Davydov}, \citenamefont
  {Belyaev}, \citenamefont {Toropov},\ and\ \citenamefont {Gil}}]{Shubina2019}%
  \BibitemOpen
  \bibfield  {author} {\bibinfo {author} {\bibfnamefont {T.~V.}\ \bibnamefont
  {Shubina}}, \bibinfo {author} {\bibfnamefont {M.}~\bibnamefont {Remškar}},
  \bibinfo {author} {\bibfnamefont {V.~Y.}\ \bibnamefont {Davydov}}, \bibinfo
  {author} {\bibfnamefont {K.~G.}\ \bibnamefont {Belyaev}}, \bibinfo {author}
  {\bibfnamefont {A.~A.}\ \bibnamefont {Toropov}},\ and\ \bibinfo {author}
  {\bibfnamefont {B.}~\bibnamefont {Gil}},\ }\bibfield  {title} {\bibinfo
  {title} {Excitonic emission in van der \uppercase{W}aals nanotubes of
  transition metal dichalcogenides},\ }\href
  {https://doi.org/10.1002/andp.201800415} {\bibfield  {journal} {\bibinfo
  {journal} {Ann. Phys.}\ }\textbf {\bibinfo {volume} {531}},\ \bibinfo {pages}
  {1800415} (\bibinfo {year} {2019})}\BibitemShut {NoStop}%
\bibitem [{\citenamefont {Gao}\ and\ \citenamefont {Xu}(2018)}]{Gao2018}%
  \BibitemOpen
  \bibfield  {author} {\bibinfo {author} {\bibfnamefont {Y.}~\bibnamefont
  {Gao}}\ and\ \bibinfo {author} {\bibfnamefont {B.}~\bibnamefont {Xu}},\
  }\bibfield  {title} {\bibinfo {title} {On the generalized thermal conductance
  characterizations of mixed one-dimensional–two-dimensional van der
  \uppercase{W}aals heterostructures and their implication for pressure
  sensors},\ }\href {https://doi.org/10.1021/acsami.8b03752} {\bibfield
  {journal} {\bibinfo  {journal} {ACS Appl. Mater. Interfaces}\ }\textbf
  {\bibinfo {volume} {10}},\ \bibinfo {pages} {14221} (\bibinfo {year}
  {2018})}\BibitemShut {NoStop}%
\bibitem [{\citenamefont {Nakanishi}\ \emph {et~al.}(2023)\citenamefont
  {Nakanishi}, \citenamefont {Furusawa}, \citenamefont {Sato}, \citenamefont
  {Tanaka}, \citenamefont {Yomogida}, \citenamefont {Yanagi}, \citenamefont
  {Zhang}, \citenamefont {Nakajo}, \citenamefont {Aoki}, \citenamefont {Kato},
  \citenamefont {Suenaga},\ and\ \citenamefont {Miyata}}]{Nakanishi2023}%
  \BibitemOpen
  \bibfield  {author} {\bibinfo {author} {\bibfnamefont {Y.}~\bibnamefont
  {Nakanishi}}, \bibinfo {author} {\bibfnamefont {S.}~\bibnamefont {Furusawa}},
  \bibinfo {author} {\bibfnamefont {Y.}~\bibnamefont {Sato}}, \bibinfo {author}
  {\bibfnamefont {T.}~\bibnamefont {Tanaka}}, \bibinfo {author} {\bibfnamefont
  {Y.}~\bibnamefont {Yomogida}}, \bibinfo {author} {\bibfnamefont
  {K.}~\bibnamefont {Yanagi}}, \bibinfo {author} {\bibfnamefont
  {W.}~\bibnamefont {Zhang}}, \bibinfo {author} {\bibfnamefont
  {H.}~\bibnamefont {Nakajo}}, \bibinfo {author} {\bibfnamefont
  {S.}~\bibnamefont {Aoki}}, \bibinfo {author} {\bibfnamefont {T.}~\bibnamefont
  {Kato}}, \bibinfo {author} {\bibfnamefont {K.}~\bibnamefont {Suenaga}},\ and\
  \bibinfo {author} {\bibfnamefont {Y.}~\bibnamefont {Miyata}},\ }\bibfield
  {title} {\bibinfo {title} {Structural diversity of single‐walled transition
  metal dichalcogenide nanotubes grown via template reaction},\ }\href
  {https://doi.org/10.1002/adma.202306631} {\bibfield  {journal} {\bibinfo
  {journal} {Adv. Mater.}\ }\textbf {\bibinfo {volume} {35}},\ \bibinfo {pages}
  {2306631} (\bibinfo {year} {2023})}\BibitemShut {NoStop}%
\bibitem [{\citenamefont {Kamaei}\ \emph {et~al.}(2020)\citenamefont {Kamaei},
  \citenamefont {Saeidi}, \citenamefont {Jazaeri}, \citenamefont {Rassekh},
  \citenamefont {Oliva}, \citenamefont {Cavalieri}, \citenamefont {Lambert},\
  and\ \citenamefont {Ionescu}}]{Kamaei2020}%
  \BibitemOpen
  \bibfield  {author} {\bibinfo {author} {\bibfnamefont {S.}~\bibnamefont
  {Kamaei}}, \bibinfo {author} {\bibfnamefont {A.}~\bibnamefont {Saeidi}},
  \bibinfo {author} {\bibfnamefont {F.}~\bibnamefont {Jazaeri}}, \bibinfo
  {author} {\bibfnamefont {A.}~\bibnamefont {Rassekh}}, \bibinfo {author}
  {\bibfnamefont {N.}~\bibnamefont {Oliva}}, \bibinfo {author} {\bibfnamefont
  {M.}~\bibnamefont {Cavalieri}}, \bibinfo {author} {\bibfnamefont
  {B.}~\bibnamefont {Lambert}},\ and\ \bibinfo {author} {\bibfnamefont {A.~M.}\
  \bibnamefont {Ionescu}},\ }\bibfield  {title} {\bibinfo {title} {An
  experimental study on mixed-dimensional 1\uppercase{D}-2\uppercase{D} van der
  \uppercase{W}aals single-walled carbon nanotube-\uppercase{WS}e$_2$
  hetero-junction},\ }\href {https://doi.org/10.1109/led.2020.2974400}
  {\bibfield  {journal} {\bibinfo  {journal} {IEEE Electron Device Lett.}\
  }\textbf {\bibinfo {volume} {41}},\ \bibinfo {pages} {645} (\bibinfo {year}
  {2020})}\BibitemShut {NoStop}%
\bibitem [{\citenamefont {Qin}\ \emph {et~al.}(2017)\citenamefont {Qin},
  \citenamefont {Shi}, \citenamefont {Ideue}, \citenamefont {Yoshida},
  \citenamefont {Zak}, \citenamefont {Tenne}, \citenamefont {Kikitsu},
  \citenamefont {Inoue}, \citenamefont {Hashizume},\ and\ \citenamefont
  {Iwasa}}]{Qin2017}%
  \BibitemOpen
  \bibfield  {author} {\bibinfo {author} {\bibfnamefont {F.}~\bibnamefont
  {Qin}}, \bibinfo {author} {\bibfnamefont {W.}~\bibnamefont {Shi}}, \bibinfo
  {author} {\bibfnamefont {T.}~\bibnamefont {Ideue}}, \bibinfo {author}
  {\bibfnamefont {M.}~\bibnamefont {Yoshida}}, \bibinfo {author} {\bibfnamefont
  {A.}~\bibnamefont {Zak}}, \bibinfo {author} {\bibfnamefont {R.}~\bibnamefont
  {Tenne}}, \bibinfo {author} {\bibfnamefont {T.}~\bibnamefont {Kikitsu}},
  \bibinfo {author} {\bibfnamefont {D.}~\bibnamefont {Inoue}}, \bibinfo
  {author} {\bibfnamefont {D.}~\bibnamefont {Hashizume}},\ and\ \bibinfo
  {author} {\bibfnamefont {Y.}~\bibnamefont {Iwasa}},\ }\bibfield  {title}
  {\bibinfo {title} {Superconductivity in a chiral nanotube},\ }\href
  {https://doi.org/10.1038/ncomms14465} {\bibfield  {journal} {\bibinfo
  {journal} {Nat. Commun.}\ }\textbf {\bibinfo {volume} {8}},\ \bibinfo {pages}
  {14465} (\bibinfo {year} {2017})}\BibitemShut {NoStop}%
\bibitem [{\citenamefont {Qin}\ \emph {et~al.}(2018)\citenamefont {Qin},
  \citenamefont {Ideue}, \citenamefont {Shi}, \citenamefont {Zhang},
  \citenamefont {Yoshida}, \citenamefont {Zak}, \citenamefont {Tenne},
  \citenamefont {Kikitsu}, \citenamefont {Inoue}, \citenamefont {Hashizume},\
  and\ \citenamefont {Iwasa}}]{Qin2018}%
  \BibitemOpen
  \bibfield  {author} {\bibinfo {author} {\bibfnamefont {F.}~\bibnamefont
  {Qin}}, \bibinfo {author} {\bibfnamefont {T.}~\bibnamefont {Ideue}}, \bibinfo
  {author} {\bibfnamefont {W.}~\bibnamefont {Shi}}, \bibinfo {author}
  {\bibfnamefont {X.-X.}\ \bibnamefont {Zhang}}, \bibinfo {author}
  {\bibfnamefont {M.}~\bibnamefont {Yoshida}}, \bibinfo {author} {\bibfnamefont
  {A.}~\bibnamefont {Zak}}, \bibinfo {author} {\bibfnamefont {R.}~\bibnamefont
  {Tenne}}, \bibinfo {author} {\bibfnamefont {T.}~\bibnamefont {Kikitsu}},
  \bibinfo {author} {\bibfnamefont {D.}~\bibnamefont {Inoue}}, \bibinfo
  {author} {\bibfnamefont {D.}~\bibnamefont {Hashizume}},\ and\ \bibinfo
  {author} {\bibfnamefont {Y.}~\bibnamefont {Iwasa}},\ }\bibfield  {title}
  {\bibinfo {title} {Diameter-dependent superconductivity in individual
  \uppercase{WS}$_2$ nanotubes},\ }\href
  {https://doi.org/10.1021/acs.nanolett.8b02647} {\bibfield  {journal}
  {\bibinfo  {journal} {Nano Lett.}\ }\textbf {\bibinfo {volume} {18}},\
  \bibinfo {pages} {6789} (\bibinfo {year} {2018})}\BibitemShut {NoStop}%
\bibitem [{\citenamefont {Zhang}\ \emph {et~al.}(2019)\citenamefont {Zhang},
  \citenamefont {Ideue}, \citenamefont {Onga}, \citenamefont {Qin},
  \citenamefont {Suzuki}, \citenamefont {Zak}, \citenamefont {Tenne},
  \citenamefont {Smet},\ and\ \citenamefont {Iwasa}}]{Zhang2019d}%
  \BibitemOpen
  \bibfield  {author} {\bibinfo {author} {\bibfnamefont {Y.~J.}\ \bibnamefont
  {Zhang}}, \bibinfo {author} {\bibfnamefont {T.}~\bibnamefont {Ideue}},
  \bibinfo {author} {\bibfnamefont {M.}~\bibnamefont {Onga}}, \bibinfo {author}
  {\bibfnamefont {F.}~\bibnamefont {Qin}}, \bibinfo {author} {\bibfnamefont
  {R.}~\bibnamefont {Suzuki}}, \bibinfo {author} {\bibfnamefont
  {A.}~\bibnamefont {Zak}}, \bibinfo {author} {\bibfnamefont {R.}~\bibnamefont
  {Tenne}}, \bibinfo {author} {\bibfnamefont {J.~H.}\ \bibnamefont {Smet}},\
  and\ \bibinfo {author} {\bibfnamefont {Y.}~\bibnamefont {Iwasa}},\ }\bibfield
   {title} {\bibinfo {title} {Enhanced intrinsic photovoltaic effect in
  tungsten disulfide nanotubes},\ }\href
  {https://doi.org/10.1038/s41586-019-1303-3} {\bibfield  {journal} {\bibinfo
  {journal} {Nature}\ }\textbf {\bibinfo {volume} {570}},\ \bibinfo {pages}
  {349} (\bibinfo {year} {2019})}\BibitemShut {NoStop}%
\bibitem [{\citenamefont {Kim}\ \emph {et~al.}(2022)\citenamefont {Kim},
  \citenamefont {Park},\ and\ \citenamefont {Kim}}]{Kim2022}%
  \BibitemOpen
  \bibfield  {author} {\bibinfo {author} {\bibfnamefont {B.}~\bibnamefont
  {Kim}}, \bibinfo {author} {\bibfnamefont {N.}~\bibnamefont {Park}},\ and\
  \bibinfo {author} {\bibfnamefont {J.}~\bibnamefont {Kim}},\ }\bibfield
  {title} {\bibinfo {title} {Giant bulk photovoltaic effect driven by the
  wall-to-wall charge shift in \uppercase{WS}$_2$ nanotubes},\ }\href
  {https://doi.org/10.1038/s41467-022-31018-8} {\bibfield  {journal} {\bibinfo
  {journal} {Nat. Commun.}\ }\textbf {\bibinfo {volume} {13}},\ \bibinfo
  {pages} {3237} (\bibinfo {year} {2022})}\BibitemShut {NoStop}%
\bibitem [{\citenamefont {Evarestov}\ \emph {et~al.}(2017)\citenamefont
  {Evarestov}, \citenamefont {Bandura}, \citenamefont {Porsev},\ and\
  \citenamefont {Kovalenko}}]{Evarestov2017}%
  \BibitemOpen
  \bibfield  {author} {\bibinfo {author} {\bibfnamefont {R.~A.}\ \bibnamefont
  {Evarestov}}, \bibinfo {author} {\bibfnamefont {A.~V.}\ \bibnamefont
  {Bandura}}, \bibinfo {author} {\bibfnamefont {V.~V.}\ \bibnamefont
  {Porsev}},\ and\ \bibinfo {author} {\bibfnamefont {A.~V.}\ \bibnamefont
  {Kovalenko}},\ }\bibfield  {title} {\bibinfo {title} {Phonon spectra,
  electronic, and thermodynamic properties of \uppercase{WS}$_2$ nanotubes},\
  }\href {https://doi.org/10.1002/jcc.24916} {\bibfield  {journal} {\bibinfo
  {journal} {J. Comput. Chem.}\ }\textbf {\bibinfo {volume} {38}},\ \bibinfo
  {pages} {2581} (\bibinfo {year} {2017})}\BibitemShut {NoStop}%
\bibitem [{\citenamefont {Zhao}\ \emph {et~al.}(2015)\citenamefont {Zhao},
  \citenamefont {Li}, \citenamefont {Duan},\ and\ \citenamefont
  {Ding}}]{Zhao2015}%
  \BibitemOpen
  \bibfield  {author} {\bibinfo {author} {\bibfnamefont {W.}~\bibnamefont
  {Zhao}}, \bibinfo {author} {\bibfnamefont {Y.}~\bibnamefont {Li}}, \bibinfo
  {author} {\bibfnamefont {W.}~\bibnamefont {Duan}},\ and\ \bibinfo {author}
  {\bibfnamefont {F.}~\bibnamefont {Ding}},\ }\bibfield  {title} {\bibinfo
  {title} {Ultra-stable small diameter hybrid transition metal dichalcogenide
  nanotubes \uppercase{X}–\uppercase{M}–\uppercase{Y} (\uppercase{X},
  \uppercase{Y} = \uppercase{S}, \uppercase{S}e, \uppercase{T}e; \uppercase{M}
  = \uppercase{M}o, \uppercase{W}, \uppercase{N}b, \uppercase{T}a): a
  computational study},\ }\href {https://doi.org/10.1039/c5nr02812d} {\bibfield
   {journal} {\bibinfo  {journal} {Nanoscale}\ }\textbf {\bibinfo {volume}
  {7}},\ \bibinfo {pages} {13586} (\bibinfo {year} {2015})}\BibitemShut
  {NoStop}%
\bibitem [{\citenamefont {Bölle}\ \emph {et~al.}(2021)\citenamefont {Bölle},
  \citenamefont {Mikkelsen}, \citenamefont {Thygesen}, \citenamefont {Vegge},\
  and\ \citenamefont {Castelli}}]{Boelle2021}%
  \BibitemOpen
  \bibfield  {author} {\bibinfo {author} {\bibfnamefont {F.~T.}\ \bibnamefont
  {Bölle}}, \bibinfo {author} {\bibfnamefont {A.~E.~G.}\ \bibnamefont
  {Mikkelsen}}, \bibinfo {author} {\bibfnamefont {K.~S.}\ \bibnamefont
  {Thygesen}}, \bibinfo {author} {\bibfnamefont {T.}~\bibnamefont {Vegge}},\
  and\ \bibinfo {author} {\bibfnamefont {I.~E.}\ \bibnamefont {Castelli}},\
  }\bibfield  {title} {\bibinfo {title} {Structural and chemical mechanisms
  governing stability of inorganic \uppercase{J}anus nanotubes},\ }\href
  {https://doi.org/10.1038/s41524-021-00505-9} {\bibfield  {journal} {\bibinfo
  {journal} {npj Comput. Mater.}\ }\textbf {\bibinfo {volume} {7}},\ \bibinfo
  {pages} {41} (\bibinfo {year} {2021})}\BibitemShut {NoStop}%
\bibitem [{\citenamefont {Bhardwaj}\ \emph {et~al.}(2021)\citenamefont
  {Bhardwaj}, \citenamefont {Sharma},\ and\ \citenamefont
  {Suryanarayana}}]{Bhardwaj2021}%
  \BibitemOpen
  \bibfield  {author} {\bibinfo {author} {\bibfnamefont {A.}~\bibnamefont
  {Bhardwaj}}, \bibinfo {author} {\bibfnamefont {A.}~\bibnamefont {Sharma}},\
  and\ \bibinfo {author} {\bibfnamefont {P.}~\bibnamefont {Suryanarayana}},\
  }\bibfield  {title} {\bibinfo {title} {Torsional strain engineering of
  transition metal dichalcogenide nanotubes: an ab initio study},\ }\href
  {https://doi.org/10.1088/1361-6528/ac1a90} {\bibfield  {journal} {\bibinfo
  {journal} {Proc. Spie.}\ }\textbf {\bibinfo {volume} {32}},\ \bibinfo {pages}
  {47LT01} (\bibinfo {year} {2021})}\BibitemShut {NoStop}%
\bibitem [{SM()}]{SM}%
  \BibitemOpen
  \href@noop {} {\bibinfo  {journal} {See supplemental material}\ }\BibitemShut
  {NoStop}%
\bibitem [{\citenamefont {Mermin}(1979)}]{Mermin1979}%
  \BibitemOpen
\bibfield  {journal} {  }\bibfield  {author} {\bibinfo {author} {\bibfnamefont
  {N.~D.}\ \bibnamefont {Mermin}},\ }\bibfield  {title} {\bibinfo {title} {The
  topological theory of defects in ordered media},\ }\href
  {https://doi.org/10.1103/revmodphys.51.591} {\bibfield  {journal} {\bibinfo
  {journal} {Rev. Mod. Phys.}\ }\textbf {\bibinfo {volume} {51}},\ \bibinfo
  {pages} {591} (\bibinfo {year} {1979})}\BibitemShut {NoStop}%
\bibitem [{\citenamefont {Verba}\ \emph {et~al.}(2018)\citenamefont {Verba},
  \citenamefont {Navas}, \citenamefont {Hierro-Rodriguez}, \citenamefont
  {Bunyaev}, \citenamefont {Ivanov}, \citenamefont {Guslienko},\ and\
  \citenamefont {Kakazei}}]{Verba2018}%
  \BibitemOpen
  \bibfield  {author} {\bibinfo {author} {\bibfnamefont {R.~V.}\ \bibnamefont
  {Verba}}, \bibinfo {author} {\bibfnamefont {D.}~\bibnamefont {Navas}},
  \bibinfo {author} {\bibfnamefont {A.}~\bibnamefont {Hierro-Rodriguez}},
  \bibinfo {author} {\bibfnamefont {S.~A.}\ \bibnamefont {Bunyaev}}, \bibinfo
  {author} {\bibfnamefont {B.~A.}\ \bibnamefont {Ivanov}}, \bibinfo {author}
  {\bibfnamefont {K.~Y.}\ \bibnamefont {Guslienko}},\ and\ \bibinfo {author}
  {\bibfnamefont {G.~N.}\ \bibnamefont {Kakazei}},\ }\bibfield  {title}
  {\bibinfo {title} {Overcoming the limits of vortex formation in magnetic
  nanodots by coupling to antidot matrix},\ }\href
  {https://doi.org/10.1103/physrevapplied.10.031002} {\bibfield  {journal}
  {\bibinfo  {journal} {Phys. Rev. Applied}\ }\textbf {\bibinfo {volume}
  {10}},\ \bibinfo {pages} {031002} (\bibinfo {year} {2018})}\BibitemShut
  {NoStop}%
\bibitem [{\citenamefont {Shibata}\ \emph {et~al.}(2006)\citenamefont
  {Shibata}, \citenamefont {Nakatani}, \citenamefont {Tatara}, \citenamefont
  {Kohno},\ and\ \citenamefont {Otani}}]{Shibata2006}%
  \BibitemOpen
  \bibfield  {author} {\bibinfo {author} {\bibfnamefont {J.}~\bibnamefont
  {Shibata}}, \bibinfo {author} {\bibfnamefont {Y.}~\bibnamefont {Nakatani}},
  \bibinfo {author} {\bibfnamefont {G.}~\bibnamefont {Tatara}}, \bibinfo
  {author} {\bibfnamefont {H.}~\bibnamefont {Kohno}},\ and\ \bibinfo {author}
  {\bibfnamefont {Y.}~\bibnamefont {Otani}},\ }\bibfield  {title} {\bibinfo
  {title} {Current-induced magnetic vortex motion by spin-transfer torque},\
  }\href {https://doi.org/10.1103/physrevb.73.020403} {\bibfield  {journal}
  {\bibinfo  {journal} {Phys. Rev. B}\ }\textbf {\bibinfo {volume} {73}},\
  \bibinfo {pages} {020403} (\bibinfo {year} {2006})}\BibitemShut {NoStop}%
\bibitem [{\citenamefont {Wong}\ \emph {et~al.}(2019)\citenamefont {Wong},
  \citenamefont {Zhang}, \citenamefont {Bussolotti}, \citenamefont {Yin},
  \citenamefont {Herng}, \citenamefont {Zhang}, \citenamefont {Huang},
  \citenamefont {Vinai}, \citenamefont {Krishnamurthi}, \citenamefont
  {Bukhvalov}, \citenamefont {Zheng}, \citenamefont {Chua}, \citenamefont
  {N’Diaye}, \citenamefont {Morton}, \citenamefont {Yang}, \citenamefont
  {Ou~Yang}, \citenamefont {Torelli}, \citenamefont {Chen}, \citenamefont
  {Goh}, \citenamefont {Ding}, \citenamefont {Lin}, \citenamefont {Brocks},
  \citenamefont {de~Jong}, \citenamefont {Castro~Neto},\ and\ \citenamefont
  {Wee}}]{Wong2019}%
  \BibitemOpen
  \bibfield  {author} {\bibinfo {author} {\bibfnamefont {P.~K.~J.}\
  \bibnamefont {Wong}}, \bibinfo {author} {\bibfnamefont {W.}~\bibnamefont
  {Zhang}}, \bibinfo {author} {\bibfnamefont {F.}~\bibnamefont {Bussolotti}},
  \bibinfo {author} {\bibfnamefont {X.}~\bibnamefont {Yin}}, \bibinfo {author}
  {\bibfnamefont {T.~S.}\ \bibnamefont {Herng}}, \bibinfo {author}
  {\bibfnamefont {L.}~\bibnamefont {Zhang}}, \bibinfo {author} {\bibfnamefont
  {Y.~L.}\ \bibnamefont {Huang}}, \bibinfo {author} {\bibfnamefont
  {G.}~\bibnamefont {Vinai}}, \bibinfo {author} {\bibfnamefont
  {S.}~\bibnamefont {Krishnamurthi}}, \bibinfo {author} {\bibfnamefont {D.~W.}\
  \bibnamefont {Bukhvalov}}, \bibinfo {author} {\bibfnamefont {Y.~J.}\
  \bibnamefont {Zheng}}, \bibinfo {author} {\bibfnamefont {R.}~\bibnamefont
  {Chua}}, \bibinfo {author} {\bibfnamefont {A.~T.}\ \bibnamefont {N’Diaye}},
  \bibinfo {author} {\bibfnamefont {S.~A.}\ \bibnamefont {Morton}}, \bibinfo
  {author} {\bibfnamefont {C.}~\bibnamefont {Yang}}, \bibinfo {author}
  {\bibfnamefont {K.}~\bibnamefont {Ou~Yang}}, \bibinfo {author} {\bibfnamefont
  {P.}~\bibnamefont {Torelli}}, \bibinfo {author} {\bibfnamefont
  {W.}~\bibnamefont {Chen}}, \bibinfo {author} {\bibfnamefont {K.~E.~J.}\
  \bibnamefont {Goh}}, \bibinfo {author} {\bibfnamefont {J.}~\bibnamefont
  {Ding}}, \bibinfo {author} {\bibfnamefont {M.}~\bibnamefont {Lin}}, \bibinfo
  {author} {\bibfnamefont {G.}~\bibnamefont {Brocks}}, \bibinfo {author}
  {\bibfnamefont {M.~P.}\ \bibnamefont {de~Jong}}, \bibinfo {author}
  {\bibfnamefont {A.~H.}\ \bibnamefont {Castro~Neto}},\ and\ \bibinfo {author}
  {\bibfnamefont {A.~T.~S.}\ \bibnamefont {Wee}},\ }\bibfield  {title}
  {\bibinfo {title} {Evidence of spin frustration in a vanadium diselenide
  monolayer magnet},\ }\href {https://doi.org/10.1002/adma.201901185}
  {\bibfield  {journal} {\bibinfo  {journal} {Adv. Mater.}\ }\textbf {\bibinfo
  {volume} {31}},\ \bibinfo {pages} {1901185} (\bibinfo {year}
  {2019})}\BibitemShut {NoStop}%
\bibitem [{\citenamefont {Memarzadeh}\ \emph {et~al.}(2021)\citenamefont
  {Memarzadeh}, \citenamefont {Roknabadi}, \citenamefont {Modarresi},
  \citenamefont {Mogulkoc},\ and\ \citenamefont {Rudenko}}]{Memarzadeh2021}%
  \BibitemOpen
  \bibfield  {author} {\bibinfo {author} {\bibfnamefont {S.}~\bibnamefont
  {Memarzadeh}}, \bibinfo {author} {\bibfnamefont {M.~R.}\ \bibnamefont
  {Roknabadi}}, \bibinfo {author} {\bibfnamefont {M.}~\bibnamefont
  {Modarresi}}, \bibinfo {author} {\bibfnamefont {A.}~\bibnamefont
  {Mogulkoc}},\ and\ \bibinfo {author} {\bibfnamefont {A.~N.}\ \bibnamefont
  {Rudenko}},\ }\bibfield  {title} {\bibinfo {title} {Role of charge doping and
  strain in the stabilization of in-plane ferromagnetism in monolayer
  \uppercase{VS}e$_2$ at room temperature},\ }\href
  {https://doi.org/10.1088/2053-1583/abf626} {\bibfield  {journal} {\bibinfo
  {journal} {2D Mater.}\ }\textbf {\bibinfo {volume} {8}},\ \bibinfo {pages}
  {035022} (\bibinfo {year} {2021})}\BibitemShut {NoStop}%
\bibitem [{\citenamefont {Yu}\ \emph {et~al.}(2019)\citenamefont {Yu},
  \citenamefont {Li}, \citenamefont {Herng}, \citenamefont {Wang},
  \citenamefont {Zhao}, \citenamefont {Chi}, \citenamefont {Fu}, \citenamefont
  {Abdelwahab}, \citenamefont {Zhou}, \citenamefont {Dan}, \citenamefont
  {Chen}, \citenamefont {Chen}, \citenamefont {Li}, \citenamefont {Lu},
  \citenamefont {Pennycook}, \citenamefont {Feng}, \citenamefont {Ding},\ and\
  \citenamefont {Loh}}]{Yu2019}%
  \BibitemOpen
  \bibfield  {author} {\bibinfo {author} {\bibfnamefont {W.}~\bibnamefont
  {Yu}}, \bibinfo {author} {\bibfnamefont {J.}~\bibnamefont {Li}}, \bibinfo
  {author} {\bibfnamefont {T.~S.}\ \bibnamefont {Herng}}, \bibinfo {author}
  {\bibfnamefont {Z.}~\bibnamefont {Wang}}, \bibinfo {author} {\bibfnamefont
  {X.}~\bibnamefont {Zhao}}, \bibinfo {author} {\bibfnamefont {X.}~\bibnamefont
  {Chi}}, \bibinfo {author} {\bibfnamefont {W.}~\bibnamefont {Fu}}, \bibinfo
  {author} {\bibfnamefont {I.}~\bibnamefont {Abdelwahab}}, \bibinfo {author}
  {\bibfnamefont {J.}~\bibnamefont {Zhou}}, \bibinfo {author} {\bibfnamefont
  {J.}~\bibnamefont {Dan}}, \bibinfo {author} {\bibfnamefont {Z.}~\bibnamefont
  {Chen}}, \bibinfo {author} {\bibfnamefont {Z.}~\bibnamefont {Chen}}, \bibinfo
  {author} {\bibfnamefont {Z.}~\bibnamefont {Li}}, \bibinfo {author}
  {\bibfnamefont {J.}~\bibnamefont {Lu}}, \bibinfo {author} {\bibfnamefont
  {S.~J.}\ \bibnamefont {Pennycook}}, \bibinfo {author} {\bibfnamefont {Y.~P.}\
  \bibnamefont {Feng}}, \bibinfo {author} {\bibfnamefont {J.}~\bibnamefont
  {Ding}},\ and\ \bibinfo {author} {\bibfnamefont {K.~P.}\ \bibnamefont
  {Loh}},\ }\bibfield  {title} {\bibinfo {title} {Chemically exfoliated
  \uppercase{VS}e$_2$ monolayers with roomtemperature ferromagnetism},\ }\href
  {https://doi.org/10.1002/adma.201903779} {\bibfield  {journal} {\bibinfo
  {journal} {Adv. Mater.}\ }\textbf {\bibinfo {volume} {31}},\ \bibinfo {pages}
  {1903779} (\bibinfo {year} {2019})}\BibitemShut {NoStop}%
\bibitem [{\citenamefont {Aharoni}(2000)}]{aharoni2000introduction}%
  \BibitemOpen
  \bibfield  {author} {\bibinfo {author} {\bibfnamefont {A.}~\bibnamefont
  {Aharoni}},\ }\href@noop {} {\emph {\bibinfo {title} {Introduction to the
  Theory of Ferromagnetism}}},\ Vol.\ \bibinfo {volume} {109}\ (\bibinfo
  {publisher} {Clarendon Press},\ \bibinfo {year} {2000})\BibitemShut {NoStop}%
\bibitem [{\citenamefont {Jia}\ \emph {et~al.}(2019{\natexlab{a}})\citenamefont
  {Jia}, \citenamefont {Ma}, \citenamefont {Schäffer},\ and\ \citenamefont
  {Berakdar}}]{Jia2019}%
  \BibitemOpen
  \bibfield  {author} {\bibinfo {author} {\bibfnamefont {C.}~\bibnamefont
  {Jia}}, \bibinfo {author} {\bibfnamefont {D.}~\bibnamefont {Ma}}, \bibinfo
  {author} {\bibfnamefont {A.~F.}\ \bibnamefont {Schäffer}},\ and\ \bibinfo
  {author} {\bibfnamefont {J.}~\bibnamefont {Berakdar}},\ }\bibfield  {title}
  {\bibinfo {title} {Twisted magnon beams carrying orbital angular momentum},\
  }\href {https://doi.org/10.1038/s41467-019-10008-3} {\bibfield  {journal}
  {\bibinfo  {journal} {Nat. Commun.}\ }\textbf {\bibinfo {volume} {10}},\
  \bibinfo {pages} {2077} (\bibinfo {year} {2019}{\natexlab{a}})}\BibitemShut
  {NoStop}%
\bibitem [{\citenamefont {Jiang}\ \emph {et~al.}(2020)\citenamefont {Jiang},
  \citenamefont {Yuan}, \citenamefont {Li}, \citenamefont {Wang}, \citenamefont
  {Zhang}, \citenamefont {Cao},\ and\ \citenamefont {Yan}}]{Jiang2020}%
  \BibitemOpen
  \bibfield  {author} {\bibinfo {author} {\bibfnamefont {Y.}~\bibnamefont
  {Jiang}}, \bibinfo {author} {\bibfnamefont {H.}~\bibnamefont {Yuan}},
  \bibinfo {author} {\bibfnamefont {Z.-X.}\ \bibnamefont {Li}}, \bibinfo
  {author} {\bibfnamefont {Z.}~\bibnamefont {Wang}}, \bibinfo {author}
  {\bibfnamefont {H.}~\bibnamefont {Zhang}}, \bibinfo {author} {\bibfnamefont
  {Y.}~\bibnamefont {Cao}},\ and\ \bibinfo {author} {\bibfnamefont
  {P.}~\bibnamefont {Yan}},\ }\bibfield  {title} {\bibinfo {title} {Twisted
  magnon as a magnetic tweezer},\ }\href
  {https://doi.org/10.1103/physrevlett.124.217204} {\bibfield  {journal}
  {\bibinfo  {journal} {Phys. Rev. Lett.}\ }\textbf {\bibinfo {volume} {124}},\
  \bibinfo {pages} {217204} (\bibinfo {year} {2020})}\BibitemShut {NoStop}%
\bibitem [{\citenamefont {Chen}\ \emph {et~al.}(2020)\citenamefont {Chen},
  \citenamefont {Schäffer}, \citenamefont {Berakdar},\ and\ \citenamefont
  {Jia}}]{Chen2020c}%
  \BibitemOpen
  \bibfield  {author} {\bibinfo {author} {\bibfnamefont {M.}~\bibnamefont
  {Chen}}, \bibinfo {author} {\bibfnamefont {A.~F.}\ \bibnamefont {Schäffer}},
  \bibinfo {author} {\bibfnamefont {J.}~\bibnamefont {Berakdar}},\ and\
  \bibinfo {author} {\bibfnamefont {C.}~\bibnamefont {Jia}},\ }\bibfield
  {title} {\bibinfo {title} {Generation, electric detection, and
  orbital-angular momentum tunneling of twisted magnons},\ }\href
  {https://doi.org/10.1063/5.0005764} {\bibfield  {journal} {\bibinfo
  {journal} {Appl. Phys. Lett.}\ }\textbf {\bibinfo {volume} {116}},\ \bibinfo
  {pages} {172403} (\bibinfo {year} {2020})}\BibitemShut {NoStop}%
\bibitem [{\citenamefont {Jia}\ \emph {et~al.}(2019{\natexlab{b}})\citenamefont
  {Jia}, \citenamefont {Ma}, \citenamefont {Schäffer},\ and\ \citenamefont
  {Berakdar}}]{Jia2019a}%
  \BibitemOpen
  \bibfield  {author} {\bibinfo {author} {\bibfnamefont {C.}~\bibnamefont
  {Jia}}, \bibinfo {author} {\bibfnamefont {D.}~\bibnamefont {Ma}}, \bibinfo
  {author} {\bibfnamefont {A.~F.}\ \bibnamefont {Schäffer}},\ and\ \bibinfo
  {author} {\bibfnamefont {J.}~\bibnamefont {Berakdar}},\ }\bibfield  {title}
  {\bibinfo {title} {Twisting and tweezing the spin wave: on vortices,
  skyrmions, helical waves, and the magnonic spiral phase plate},\ }\href
  {https://doi.org/10.1088/2040-8986/ab4f8e} {\bibfield  {journal} {\bibinfo
  {journal} {J. Opt.}\ }\textbf {\bibinfo {volume} {21}},\ \bibinfo {pages}
  {124001} (\bibinfo {year} {2019}{\natexlab{b}})}\BibitemShut {NoStop}%
\bibitem [{\citenamefont {Lee}\ and\ \citenamefont {Kim}(2022)}]{Lee2022}%
  \BibitemOpen
  \bibfield  {author} {\bibinfo {author} {\bibfnamefont {S.}~\bibnamefont
  {Lee}}\ and\ \bibinfo {author} {\bibfnamefont {S.~K.}\ \bibnamefont {Kim}},\
  }\bibfield  {title} {\bibinfo {title} {Generation of magnon orbital angular
  momentum by a skyrmion-textured domain wall in a ferromagnetic nanotube},\
  }\href {https://doi.org/10.3389/fphy.2022.858614} {\bibfield  {journal}
  {\bibinfo  {journal} {Aip. Conf. Proc.}\ }\textbf {\bibinfo {volume} {10}},\
  \bibinfo {pages} {858614} (\bibinfo {year} {2022})}\BibitemShut {NoStop}%
\bibitem [{\citenamefont {Li}\ \emph {et~al.}(2022{\natexlab{b}})\citenamefont
  {Li}, \citenamefont {Wang}, \citenamefont {Cao},\ and\ \citenamefont
  {Yan}}]{Li2022c}%
  \BibitemOpen
  \bibfield  {author} {\bibinfo {author} {\bibfnamefont {Z.-X.}\ \bibnamefont
  {Li}}, \bibinfo {author} {\bibfnamefont {Z.}~\bibnamefont {Wang}}, \bibinfo
  {author} {\bibfnamefont {Y.}~\bibnamefont {Cao}},\ and\ \bibinfo {author}
  {\bibfnamefont {P.}~\bibnamefont {Yan}},\ }\bibfield  {title} {\bibinfo
  {title} {Generation of twisted magnons via spin-to-orbital angular momentum
  conversion},\ }\href {https://doi.org/10.1103/physrevb.105.174433} {\bibfield
   {journal} {\bibinfo  {journal} {Phys. Rev. B}\ }\textbf {\bibinfo {volume}
  {105}},\ \bibinfo {pages} {174433} (\bibinfo {year}
  {2022}{\natexlab{b}})}\BibitemShut {NoStop}%
\bibitem [{\citenamefont {Jia}\ \emph {et~al.}(2021)\citenamefont {Jia},
  \citenamefont {Chen}, \citenamefont {Schäffer},\ and\ \citenamefont
  {Berakdar}}]{Jia2021}%
  \BibitemOpen
  \bibfield  {author} {\bibinfo {author} {\bibfnamefont {C.}~\bibnamefont
  {Jia}}, \bibinfo {author} {\bibfnamefont {M.}~\bibnamefont {Chen}}, \bibinfo
  {author} {\bibfnamefont {A.~F.}\ \bibnamefont {Schäffer}},\ and\ \bibinfo
  {author} {\bibfnamefont {J.}~\bibnamefont {Berakdar}},\ }\bibfield  {title}
  {\bibinfo {title} {Chiral logic computing with twisted antiferromagnetic
  magnon modes},\ }\href {https://doi.org/10.1038/s41524-021-00570-0}
  {\bibfield  {journal} {\bibinfo  {journal} {npj Comput. Mater.}\ }\textbf
  {\bibinfo {volume} {7}},\ \bibinfo {pages} {101} (\bibinfo {year}
  {2021})}\BibitemShut {NoStop}%
\bibitem [{\citenamefont {Wang}\ \emph {et~al.}(2022)\citenamefont {Wang},
  \citenamefont {Yuan}, \citenamefont {Cao},\ and\ \citenamefont
  {Yan}}]{Wang2022a}%
  \BibitemOpen
  \bibfield  {author} {\bibinfo {author} {\bibfnamefont {Z.}~\bibnamefont
  {Wang}}, \bibinfo {author} {\bibfnamefont {H.}~\bibnamefont {Yuan}}, \bibinfo
  {author} {\bibfnamefont {Y.}~\bibnamefont {Cao}},\ and\ \bibinfo {author}
  {\bibfnamefont {P.}~\bibnamefont {Yan}},\ }\bibfield  {title} {\bibinfo
  {title} {Twisted magnon frequency comb and penrose superradiance},\ }\href
  {https://doi.org/10.1103/physrevlett.129.107203} {\bibfield  {journal}
  {\bibinfo  {journal} {Phys. Rev. Lett.}\ }\textbf {\bibinfo {volume} {129}},\
  \bibinfo {pages} {107203} (\bibinfo {year} {2022})}\BibitemShut {NoStop}%
\bibitem [{\citenamefont {Gilbert}(2004)}]{Gilbert2004}%
  \BibitemOpen
  \bibfield  {author} {\bibinfo {author} {\bibfnamefont {T.}~\bibnamefont
  {Gilbert}},\ }\bibfield  {title} {\bibinfo {title} {Classics in magnetics a
  phenomenological theory of damping in ferromagnetic materials},\ }\href
  {https://doi.org/10.1109/tmag.2004.836740} {\bibfield  {journal} {\bibinfo
  {journal} {IEEE Trans. Magn.}\ }\textbf {\bibinfo {volume} {40}},\ \bibinfo
  {pages} {3443} (\bibinfo {year} {2004})}\BibitemShut {NoStop}%
\end{thebibliography}%

\end{document}